\documentstyle[12pt]{article}
\catcode`\@=11
\def\numberbysection{\@addtoreset{equation}{section}
        \def\theequation{\thesection.\arabic{equation}}}
\numberbysection

\begin{document}

\newlength{\lno} \lno0.5cm \newlength{\len} \len=\textwidth%
\addtolength{\len}{-\lno}

\setcounter{page}{0}

\baselineskip7mm \renewcommand{\thefootnote}{\fnsymbol{footnote}} \newpage %
\setcounter{page}{0}

\begin{titlepage}     
\vspace{1.5cm}
\begin{center}
{\Large\bf Reflection K-Matrices for 19-Vertex Models}\\
\vspace{1cm}
{\large A. Lima-Santos }\footnote{e-mail: dals@power.ufscar.br} \\
\vspace{1cm}
{\large \em Universidade Federal de S\~ao Carlos, Departamento de F\'{\i}sica \\
Caixa Postal 676, CEP 13569-905~~S\~ao Carlos, Brasil}\\
\end{center}
\vspace{2.5cm}

\begin{abstract}
We derive and classify all regular solutions of the boundary Yang-Baxter
equation for 19-vertex models known as Zamolodchikov-Fateev or $A_{1}^{(1)}$
model, Izergin-Korepin or $A_{2}^{(2)}$ model, $sl(2|1)$ model and the $%
osp(2|1)$ model. 
We find that there is a general solution for the $A_{1}^{(1)}$ and $sl(2|1)$ models. 
In both models it is a complete K-matrix with three free parameters. For the $A_{2}^{(2)}$
and $osp(2|1)$ models we  find  three general solutions, being two complete 
reflection K-matrices solutions and one incomplete reflection K-matrix solution with  
some null entries. In both models these solutions have  two free parameters. 
 Integrable spin-$1$ Hamiltonians with general boundary interactions are also
presented. 
Several reduced solutions from these general solutions are presented in the appendices.
\end{abstract}
\vfill
\begin{center}
\small{\today}
\end{center}
\end{titlepage}

\baselineskip6mm

\newpage{}

\section{{}Introduction}

In the study of the two-dimensional integrable systems of quantum field
theories and statistical physics the Yang-Baxter ({\small YB}) equation
plays essential roles in establishing the integrability and in solving the
models.

Each solution of the {\small YB} equation can be treated as a factorized $S$%
-matrix\ in some $1$+$1$-dimensional field theory \cite{Karowski, Zamo0}. In
statistical physics the {\small YB} solution is treated as a vertex weight
matrix of an exactly solvable statistical model on a plane lattice.

The central object in the theory of integrable models \cite{Baxter, Faddeev}
is the $R$-matrix $R(u)$, where $u$ is the spectral parameter. It acts on
the tensor product $V^{1}\otimes V^{2}$ for a given vector space $V$ and
satisfy a special system of algebraic functional equations, the {\small YB}
equation 
\begin{equation}
R_{12}(u)R_{13}(u+v)R_{23}(v)=R_{23}(v)R_{13}(u+v)R_{12}(u),  \label{int.1}
\end{equation}
in $V^{1}\otimes V^{2}\otimes V^{3}$, where $R_{12}=R\otimes {\bf 1}$, $%
R_{23}={\bf 1}\otimes R$, etc.

An $R$ matrix is said to be regular if has the property $R(0)=P$, where $P$
is the permutation matrix in $V^{1}\otimes V^{2}$: $P(\left| \alpha
\right\rangle \otimes \left| \beta \right\rangle )=\left| \beta
\right\rangle \otimes \left| \alpha \right\rangle $ for $\left| \alpha
\right\rangle ,\left| \beta \right\rangle \in V$. To every regular $R$%
-matrix there corresponds a periodic integrable quantum spin chain, with
Hamiltonian $H$ given by 
\begin{equation}
H=\sum_{k=1}^{N-1}H_{k,k+1}+H_{N,1},  \label{int.2}
\end{equation}
where the two-site Hamiltonian $H_{k,k+1}$ is given by 
\begin{equation}
H_{k,k+1}=P\left. \frac{d}{du}R(u)\right| _{u=0}.  \label{int.3}
\end{equation}
The right hand side of (\ref{int.3}) is acting on the quantum spaces at
sites $k$ and $k+1$.

An $R$-matrix is said to be quasi-classical if it depends on an additional
parameter $\eta $ (playing the role of the Planck constant), so that for
small $\eta $%
\begin{equation}
R(u,\eta )=1+2\eta r(u)+{\large O}(\eta ^{2})  \label{int.4}
\end{equation}
The quantity $r(u)$ is called a classical $r$-matrix.

Many regular quasi-classical solutions are related to the simple Lie (super)
algebra \cite{Kulish1,Bazhanov1,Jimbo,Bazhanov2}. They are classified by
irreducible representations of the Lie algebra in elliptic, trigonometric
and rational, corresponding to the three types of function of the spectral
parameter $u$.

Recently there has been a lot of efforts in introducing boundaries into
integrable systems for possible applications to condensed matter physics and
statistical system with non-periodic boundary conditions. The boundaries
entail new physical quantities called reflection matrices which depend on
the boundary properties.

By considering systems on a finite interval with independent boundary
conditions at each end, we have to introduce reflection matrices to describe
such boundary conditions. Integrable models with boundaries can be
constructed out of a pair of reflection $K$-matrices $K_{\pm }(u)$ in
addition to the {\small YB} $R$-matrix. $K_{+}(u)$ and $K_{-}(u)$ describe
the effects of the presence of boundaries at the left and the right ends,
respectively.

Integrability of open chains in the framework of the quantum inverse
scattering method was pioneered by Sklyanin relying on previous results of
Cherednik \cite{Cherednik}. In reference \cite{Sklyanin}, Sklyanin has used
his formalism to solve, via algebraic Bethe ansatz, the open spin-$1/2$
chain with diagonal boundary terms. This \ model had already been solved via
coordinate Bethe ansatz by Alcaraz {\it et al} \cite{Alcaraz}.

The Sklyanin original formalism was extended to more general systems by
Mezincescu and Nepomechie in \cite{Mezincescu1}, where is assumed that for a
regular $R$ matrix satisfying the following properties 
\begin{eqnarray}
{\em PT-symmetry} &:&P_{12}R_{12}(u)P_{12}=R_{21}(u),  \nonumber \\
{\em unitarity} &:&R_{12}(u)R_{21}(-u)\propto 1,  \nonumber \\
{\em crossing\ unitarity} &:&R_{12}(u)=(U\otimes 1)R_{12}^{{\rm t}%
_{2}}(-u-\rho )(U\otimes 1)^{-1},  \label{int.5}
\end{eqnarray}
one can derive an integrable open chain Hamiltonian 
\begin{equation}
H=\sum_{k=1}^{N-1}H_{k,k+1}+\frac{1}{2}(\left. \frac{dK_{-}(u)}{du}\right|
_{u=0}\otimes 1)+\frac{{\rm tr}_{0}\stackrel{0}{K}_{+}(0)H_{N,0}}{{\rm tr}%
K_{+}(0)},  \label{int.6}
\end{equation}
where $H_{k,k+1}$ is given by (\ref{int.3}) and $K_{-}(u)$ is the reflection
matrix which satisfy the right boundary {\small YB} equation, also known as
the right reflection equation ({\small RE}) 
\[
R_{12}(u-v)(K_{-}(u)\otimes 1)R_{21}(u+v)(1\otimes K_{-}(v))= 
\]
\begin{equation}
(1\otimes K_{-}(v))R_{12}(u+v)(K_{-}(u)\otimes 1)R_{21}(u-v)  \label{int.7}
\end{equation}
and $K_{+}(u)$ is the reflection matrix which satisfy an left {\small RE.}

Given a solution $K_{-}(u)$ of (\ref{int.7}), one can show that the
corresponding quantity 
\begin{equation}
K_{+}(u)=K_{-}^{{\rm t}}(-u-\rho )M,\qquad M=U^{{\rm t}}U=M^{{\rm t}},
\label{int.8}
\end{equation}
satisfy the left {\small RE}. \ Here $\rho $ is a crossing parameter and $U$
is a crossing matrix both being specific to each model \cite{Bazhanov1,
Bazhanov2}. ${\rm t}_{i}$ stands for the transposition taken in the $i$-{\it %
th} space and {\rm tr}$_{0}$ is the trace taken in the auxiliary space.

Although a careful analysis within the framework of algebraic structures was
carried out in \cite{Kulish2,Kulish3,Hlavary}, unlike $R$-matrix solutions
there is no direct relation of the quasi-classical $K$-matrix solutions to
the Lie (super) algebra theory. In particular, many of them do not depend on
the quasiclassical parameter $\eta $.

In field theory, attention is focused on the boundary $S$-matrix and recent
field theoretical applications of the {\small RE} can be found in \cite
{Cherednik2,Zamo, Konno, Fring}. In statistical mechanics, the emphasis has
been on deriving solutions of the {\small RE} and the investigation of
various surface critical phenomena, both at and away from criticality \cite
{Batch}. In condensed matter physics, the Kondo problem with integrable
boundary impurities has been studied by means of \ the boundary graded
quantum inverse scattering method \cite{Links}.

Due to the significance of the {\small RE}, a lot of work has been directed
to the study of their solutions \cite{Cherednik,deVega,Mezincescu2,Inami}.
There is a classification of reflection solutions for two-component systems 
\cite{Liu}. Batchelor {\it at al }\cite{Batchelor} have derived diagonal
solutions of the {\small RE} for the face and vertex models related to
affine Lie algebras. Complete nondiagonal reflection matrices of face-type
models as {\small SOS}, {\small RSOS} and the hard hexagonal model were
recently derived by Ahn and You \cite{Ahn}. Diagonal solutions for fused
high spin models were presented by Abad and Rios \cite{Abad}.

In this paper we consider the $19$-vertex models known as
Zamolodchikov-Fateev or $A_{1}^{(1)}$ model, Izergin-Korepin or $A_{2}^{(2)}$
model , the $sl(2|1)$ model and the $osp(2|1)$ model. These vertex models
have a common algebraic structure which permits us to solve their {\small RE}
in an unified way. General regular solutions are derived and classified into
two categories: Type-I solution, which is the complete reflection $K$-matrix
of each $19$-vertex model, and Type-II solution, which is an additional
solution for the {\small IK} model and the $osp(2|1)$ model. These solutions
were computed in a direct way by solving functional equations.

The paper is organized as follows. In section $2$ we present a procedure to
derive general solutions of the {\small RE} for these models. In section $3$
we carry out the procedure explicitly for the {\small ZF} model. Here one
can see that our result is in agreement with the previous result derived by
Inami {\it et al} \cite{Inami}. In section $4$ the type-I and type-II
general solutions are calculated for the {\small IK} model. In sections $5$
and $6$ we present the regular reflection $K$-matrices for the graded $%
sl(2|1)$ and $osp(2|1)$ models, respectively. In section $7$ the
corresponding integrable Hamiltonians with general boundary interactions are
also presented. Our conclusions are presented in the last section. Finally,
we reserved four appendices to give a sub-classification for the reduced
solutions which are derived from the general ones as special limits.{}

\section{General Solutions}

Along this paper we will work with both graded and non-graded models.
Therefore we need to recall some informations about the graded formulation.

Let $V=V_{0}\oplus V_{1}$ be a $Z_{2}$-graded vector space where $0$ and $1$
denote the even and odd parts respectively. Multiplication rules in the
graded tensor product space $V\stackrel{s}{\otimes }V$ differ from the
ordinary ones by the appearance of additional signs. The components of a
linear operator $A\stackrel{s}{\otimes }B\in V\stackrel{s}{\otimes }V$
result in matrix elements of the form 
\begin{equation}
(A\stackrel{s}{\otimes }B)_{\alpha \beta }^{\gamma \delta }=(-)^{p(\beta
)(p(\alpha )+p(\gamma ))}\ A_{\alpha \gamma }B_{\beta \delta }.
\end{equation}
The action of the graded permutation operator $P^{{\rm g}}$ on the vector $%
\left| \alpha \right\rangle \stackrel{s}{\otimes }\left| \beta \right\rangle
\in V\stackrel{s}{\otimes }V$ is defined by 
\begin{equation}
P^{{\rm g}}\ \left| \alpha \right\rangle \stackrel{s}{\otimes }\left| \beta
\right\rangle =(-)^{p(\alpha )p(\beta )}\left| \beta \right\rangle \stackrel{%
s}{\otimes }\left| \alpha \right\rangle \Longrightarrow (P^{{\rm g}%
})_{\alpha \beta }^{\gamma \delta }=(-)^{p(\alpha )p(\beta )}\delta _{\alpha
\delta }\ \delta _{\beta \gamma }.
\end{equation}
The super-transposition ${\rm st}$ and the super-trace {\rm str} are defined
by 
\begin{equation}
\left( A^{{\rm st}}\right) _{\alpha \beta }=(-)^{(p(\alpha )+p(\beta
))p(\alpha )}A_{\beta \alpha },\quad {\rm str}A=\sum_{\alpha }(-)^{p(\alpha
)}A_{\alpha \alpha }.
\end{equation}
where $p(\alpha )=1\ (0)$ if $\left| \alpha \right\rangle $ is an odd (even)
element.

Taking into account this new formulation, the equations (\ref{int.1}) and (%
\ref{int.7}) are now named graded {\small YB} equation and boundary graded 
{\small YB} equation, respectively.

The {\small YB} solution for the models which we are going to treat in this
paper has a common form given by the following $R$-matrix 
\begin{equation}
R(u)=\left( 
\begin{array}{lllllllll}
x_{1} & 0 & \ 0 & 0 & \ 0 & 0 & \ 0 & 0 & 0 \\ 
0 & x_{2} & \ 0 & x_{5} & \ 0 & 0 & \ 0 & 0 & 0 \\ 
0 & 0 & \ x_{3} & 0 & \ x_{6} & 0 & \ x_{7} & 0 & 0 \\ 
0 & y_{5} & \ 0 & x_{2} & \ 0 & 0 & 0 & 0 & 0 \\ 
0 & 0 & y_{6} & 0 & x_{4} & 0 & x_{6} & 0 & 0 \\ 
0 & 0 & \ 0 & 0 & \ 0 & x_{2} & \ 0 & x_{5} & 0 \\ 
0 & 0 & \ y_{7} & 0 & \ y_{6} & 0 & \ x_{3} & 0 & 0 \\ 
0 & 0 & \ 0 & 0 & \ 0 & y_{5} & \ 0 & x_{2} & 0 \\ 
0 & 0 & \ 0 & 0 & \ 0 & 0 & \ 0 & 0 & x_{1}
\end{array}
\right) ,  \label{eq2.1}
\end{equation}
where the non-zero entries for each model will be presented in the next
sections. In the graded cases we choose a common parity assignments: $%
p(1)=p(3)=0$, $p(2)=1$.

For regular solutions $K_{-}(u)$ of the right {\small RE} we can choose the
following normalization 
\begin{equation}
K_{-}(u)=\left( 
\begin{array}{ccc}
k_{11}(u) & k_{12}(u) & k_{13}(u) \\ 
k_{21}(u) & 1 & k_{23}(u) \\ 
k_{31}(u) & k_{32}(u) & k_{33}(u)
\end{array}
\right)  \label{eq2.2}
\end{equation}
with 
\begin{equation}
k_{ij}(0)=0\quad {\rm for}\quad i\neq j\quad {\rm and}\quad
k_{ii}(0)=1,\quad i,j=1,2,3  \label{eq2.3}
\end{equation}

Substituting (\ref{eq2.1}) and (\ref{eq2.2}) into (\ref{int.7}) and into its
graded version, we have in both cases $81$ functional equations for the $%
k_{ij}$ elements, many of which are dependent. In order to solve them, we
shall proceed in the following way. First we consider the $(i,j)$ component
of the matrix equation (\ref{int.7}). By differentiating it with respect to $%
v$ and taking $v=0$, we will get algebraic equations involving the single
variable $u$ and eight parameters 
\begin{equation}
\beta _{ij}=\frac{dk_{ij}(v)}{dv}|_{v=0}\qquad i,j=1,2,3.\qquad (\beta
_{22}=0)  \label{eq2.4}
\end{equation}
Second, these algebraic equations are denoted by $E[i,j]=0$ and collected
into $25$ blocks $B[i,j]$ defined by 
\begin{eqnarray}
B[i,j] &=&\{E[i,j]=0,\ E[j,i]=0,\ E[10-i,10-j]=0,\ E[10-j,10-i]=0\} 
\nonumber \\
i &=&1,...,5\qquad {\rm and}\qquad j=i,...,10-i.  \label{eq2.5}
\end{eqnarray}
For a given block $B[i,j]$, the equation $E[10-j,10-i]=0$ can be obtained
from the equation $E[i,j]=0$ by interchanging 
\begin{eqnarray}
k_{ij} &\longleftrightarrow &k_{4-j,4-i},\quad \beta
_{ij}\longleftrightarrow \beta _{4-j,4-i},\quad x_{i}\longleftrightarrow
y_{i}\ \quad {\rm for}\quad i=5,7  \nonumber \\
{\rm and}\quad x_{6} &\longleftrightarrow &\epsilon y_{6}\ ,  \label{eq2.6}
\end{eqnarray}
and the equation $E[j,i]=0$ is obtained from the equation $E[i,j]=0$ by the
interchanging 
\begin{equation}
k_{ij}\longleftrightarrow k_{ji},\quad \beta _{ij}\longleftrightarrow \beta
_{ji},\quad x_{6}\longleftrightarrow \epsilon x_{6}\ ,\quad
y_{6}\longleftrightarrow \epsilon y_{6}\ ,\quad   \label{eq2.7}
\end{equation}
where $\epsilon $ makes the difference among graded and non-graded models: $%
\epsilon =1$ for the {\small ZF} and {\small IK} models and $\epsilon =-1$ \
for the $sl(2|1)$ and $osp(2|1)$ models.

In order to give an example, let us consider the block $B[2,8]$ where we
have the two simplest equations: 
\begin{eqnarray}
E[2,8] &=&[\beta _{12}x_{2}y_{6}-\beta _{23}x_{3}y_{5}]k_{12}-[\epsilon
\beta _{23}x_{2}x_{6}-\beta _{12}x_{3}x_{5}]k_{23}=0,  \nonumber \\
E[8,2] &=&[\epsilon \beta _{21}x_{2}y_{6}-\beta
_{32}x_{3}y_{5}]k_{21}-[\beta _{32}x_{2}x_{6}-\beta _{21}x_{3}x_{5}]k_{32}=0.
\label{eq2.8}
\end{eqnarray}
Note that we can use (\ref{eq2.7}) to write the equation $E[8,2]=0$ from \
the equation $E[2,8]=0$. Therefore, using the interchanging rules (\ref
{eq2.6}) and (\ref{eq2.7}) for each block $B[i,j]$ we only needed to look at
the equation $E[i,j]=0$, which will itself be identified with the block

Combining the two equations of the block $B[2,8]$ with the four equations of
the block 
\begin{equation}
B[1,6]=\beta _{13}x_{2}(x_{1}-x_{3})k_{12}-\epsilon \beta
_{13}x_{5}x_{6}k_{23}-[\beta _{12}x_{2}x_{5}-\epsilon \beta
_{23}x_{1}x_{6}]k_{13},  \label{eq2.9}
\end{equation}
and with the four equations of the block 
\begin{equation}
B[1,8]=[x_{2}^{2}-x_{1}x_{3}]\beta _{12}k_{13}+\beta
_{13}x_{3}y_{5}k_{12}+\epsilon \beta _{13}x_{2}x_{6}k_{23},  \label{eq2.10}
\end{equation}
we will get the relations 
\begin{eqnarray}
k_{12} &=&\frac{\beta _{12}x_{3}x_{5}-\epsilon \beta _{23}x_{2}x_{6}}{%
x_{3}(x_{1}-x_{3})+\epsilon x_{6}y_{6}}\frac{k_{13}}{\beta _{13}},\quad
k_{23}=\frac{\beta _{23}x_{3}y_{5}-\beta _{12}x_{2}y_{6}}{%
x_{3}(x_{1}-x_{3})+\epsilon x_{6}y_{6}}\frac{k_{13}}{\beta _{13}},
\label{eq2.11} \\
k_{21} &=&\frac{\beta _{21}x_{3}x_{5}-\beta _{32}x_{2}x_{6}}{%
x_{3}(x_{1}-x_{3})+\epsilon x_{6}y_{6}}\frac{k_{31}}{\beta _{31}},\quad
k_{32}=\frac{\beta _{32}x_{3}y_{5}-\epsilon \beta _{21}x_{2}y_{6}}{%
x_{3}(x_{1}-x_{3})+\epsilon x_{6}y_{6}}\frac{k_{31}}{\beta _{31}}.
\label{eq2.12}
\end{eqnarray}
Here we have used the following identity 
\begin{equation}
x_{3}(x_{1}^{2}+x_{2}^{2})+\epsilon
x_{1}x_{6}y_{6}=x_{1}(x_{2}^{2}+x_{3}^{2})+x_{3}x_{5}y_{5}.  \label{eq2.13}
\end{equation}
which is holds for all above-mentioned $19$-vertex models.

Next we consider the nine equations of the blocks $B[i,i]$%
\begin{eqnarray}
B[1,1] &=&x_{1}x_{5}(\beta _{21}k_{12}-\beta _{12}k_{21})+x_{1}x_{7}(\beta
_{31}k_{13}-\beta _{13}k_{31}),  \nonumber \\
B[2,2] &=&x_{1}x_{5}(\beta _{12}k_{21}-\beta _{21}k_{12})+x_{5}x_{7}(\beta
_{32}k_{23}-\beta _{23}k_{32})  \nonumber \\
&&+x_{2}x_{6}(\beta _{32}k_{12}-\epsilon \beta _{23}k_{21}),  \nonumber \\
B[3,3] &=&x_{1}x_{7}(\beta _{13}k_{31}-\beta _{31}k_{13})-x_{5}x_{7}(\beta
_{32}k_{23}-\beta _{23}k_{32})  \nonumber \\
&&+x_{2}x_{6}(\epsilon \beta _{23}k_{21}-\beta _{32}k_{12}),  \nonumber \\
B[4,4] &=&\epsilon x_{4}y_{5}(\beta _{21}k_{12}-\beta
_{12}k_{21})-x_{5}y_{5}(\beta _{13}k_{31}-\beta _{31}k_{13})  \nonumber \\
&&+x_{2}x_{6}(\epsilon \beta _{21}k_{23}-\beta _{12}k_{32}),  \nonumber \\
B[5,5] &=&-\epsilon x_{4}y_{5}(\beta _{21}k_{12}-\beta _{12}k_{21})+\epsilon
x_{4}x_{5}(\beta _{32}k_{23}-\beta _{23}k_{32})  \nonumber \\
&&-x_{2}x_{6}(\epsilon \beta _{23}k_{21}-\beta _{32}k_{12})+x_{2}y_{6}(\beta
_{32}k_{12}-\epsilon \beta _{23}k_{21}),  \label{eq2.14}
\end{eqnarray}
By direct inspection one can see that these equations are solved by the
relations 
\begin{equation}
\beta _{ij}k_{ji}(u)=\beta _{ji}k_{ij}(u)\qquad i\neq j  \label{eq2.15}
\end{equation}
provided that 
\begin{equation}
\beta _{12}\beta _{32}=\epsilon \beta _{21}\beta _{23}  \label{eq2.16}
\end{equation}

Substituting (\ref{eq2.15}) and (\ref{eq2.16}) into the remaining blocks we
verify that the equations $E[i,j]=0$ and $E[j,i]=0$ give the same result
provided that 
\begin{equation}
\beta _{13}\beta _{21}^{2}=\epsilon \beta _{31}\beta _{12}^{2}
\label{eq2.17}
\end{equation}
Equation (\ref{eq2.15}) is invariant under the interchange rules (\ref{eq2.6}%
) and (\ref{eq2.7}) while equation (\ref{eq2.17}) has an interchanged form
involving $\beta _{23}$ and $\beta _{32}$.

At this stage we still have to solve the equations of $16$ blocks $B[i,j]$
involving three functions $k_{13}(u)$ , $k_{11}(u)$ and $k_{33}(u)$ and the
six parameters $\beta _{12},\beta _{13},\beta _{12,}\beta _{23},\beta _{11}$
and $\beta _{33}$.

Next, from each equation of the blocks $B[i,j]$ we express $k_{13}(u)$ in
terms of $k_{11}(u)$ and $k_{33}(u)$. Equating these results we will get
constraint equations for the parameters $\beta _{ij}$. This procedure will
leave us with many constraint equations. For instance, from the blocks $%
B[1,4]$ and $B[2,6]$ we compare the results for $k_{13}(u)$ to get the
following equation: 
\[
\epsilon \beta _{12}\beta _{21}\beta _{13}^{2}x_{1}x_{6}-\epsilon \beta
_{21}\beta _{13}^{2}[\Gamma _{1}x_{6}y_{5}+\Gamma _{2}x_{2}x_{7}]-2\beta
_{12}\beta _{13}\Gamma _{1}w(x_{1},x_{2}) 
\]
\begin{equation}
+\beta _{12}^{2}\beta _{23}\Gamma _{1}x_{2}(\epsilon x_{4}-x_{3})+\beta
_{12}^{2}\beta _{13}[\beta _{33}x_{2}x_{5}-2w(x_{2},x_{5})]=0.
\label{eq2.22}
\end{equation}
Using the interchange rule (\ref{eq2.6}) \ and the relations (\ref{eq2.16})
and (\ref{eq2.17}) we get from (\ref{eq2.22}) another constraint equation

\[
\epsilon \beta _{21}\beta _{23}\beta _{13}^{2}x_{1}y_{6}-\beta _{21}\beta
_{13}^{2}[\Gamma _{1}x_{2}y_{7}+\epsilon \Gamma _{2}x_{5}y_{6}]-2\beta
_{12}\beta _{13}\Gamma _{2}w(x_{1},x_{2}) 
\]
\begin{equation}
+\beta _{12}^{2}\beta _{23}\Gamma _{2}x_{2}(\epsilon x_{4}-x_{3})+\beta
_{12}\beta _{13}\beta _{23}[\beta _{11}x_{2}y_{5}-2w(x_{2},y_{5})]\ =0.
\label{eq2.23}
\end{equation}
Equations (\ref{eq2.22}) and (\ref{eq2.23}) are enough for our purpose. They
involve the six remaining parameters $\beta _{ij}$ and must be valid for all
values of $u$. Their solutions are model dependent and solve all other
constraint equations.

The procedure to solve these equations is the following: First we write (\ref
{eq2.22}) in a factored form $F_{1}(u,\eta )F_{2}(u,\eta )=0$ such that $%
F_{1}(u=0,\eta )=0$ and $F_{2}(u=0,\eta )\neq 0$. Imposing that $%
F_{2}(0,\eta )=0$ we can find, for instance, $\beta _{21}$ in terms of $%
\beta _{12},\beta _{13,}\beta _{23}$ and $\beta _{33}$. Substituting the
value obtained for $\beta _{21}$ into $F_{2}(u,\eta )$ we have two
possibilities. When is $F_{2}(u,\eta )=0$ it is a solution for (\ref{eq2.22}%
) with four free parameters. But if  $F_{2}(u,\eta )\neq 0$ it must be
rewrite in another factored form $F_{3}(u,\eta )F_{4}(u,\eta )=0$ such that $%
F_{3}(u=0,\eta )=0$ and $F_{4}(u=0,\eta )\neq 0$. The condition $%
F_{4}(0,\eta )=0$ can be used to find, for instance, the value of $\beta
_{33}$ in terms of $\beta _{12},\beta _{13}$ and $\beta _{23}$. Substituting 
$\beta _{33}$ into $F_{4}(u,\eta )$ we proceed as before and so on until
solving (\ref{eq2.22}) for any value of $u$.

For the {\small ZF} and the $sl(2|1)$ models we will get their parameters at
the second step of this procedure. It means that we have a general solution
with three free parameters. However, for the {\small IK} and $osp(2|1)$
models we will need an additional step that will supply us with the
following equation 
\begin{equation}
F_{5}(u,\eta )[\beta _{12}^{2}+\epsilon {\rm e}^{4\epsilon \eta }\beta
_{23}^{2}]=0,  \label{eq2.24}
\end{equation}
and its interchanged form 
\begin{equation}
G_{5}(u,\eta )[\beta _{21}^{2}+\epsilon {\rm e}^{4\epsilon \eta }\beta
_{32}^{2}]=0.
\end{equation}
Here we observe that the equations $\beta _{12}^{2}+\epsilon {\rm e}%
^{4\epsilon \eta }\beta _{23}^{2}=0$ and $\beta _{21}^{2}+\epsilon {\rm e}%
^{4\epsilon \eta }\beta _{32}^{2}=0$ will give us two general solutions with
two free parameters. Nevertheless, a third solution with $\beta _{12}=\beta
_{21}=\beta _{23}=\beta _{32}=0$ must also be considered. It is, by
construction, a new solution which also has two free parameters. In that
way, we will have two types of solutions: Type-I defined when all parameters
are non-vanishing and Type-II defined when the parameters $\beta _{12},\beta
_{21},\beta _{23}$ and $\beta _{32}$ vanish.

\subsection{Type-I Solution}

Taking into account the values of the parameters $\beta _{ij}\neq 0$ which
satisfy the constraint equations (\ref{eq2.22}) and (\ref{eq2.23}) we can
choose \ the blocks $B[1,4]$ and $B[2,4]$ to write expressions for $k_{13}$, 
$k_{11}$ and $k_{33}$. Moreover, choosing $\beta _{32}$ and $\beta _{31}$ as
the parameters fixed by the relations (\ref{eq2.16}) and (\ref{eq2.17})
respectively, we can also recall the relations (\ref{eq2.11}) and (\ref
{eq2.12}) to write the type-I solution. It is a complete reflection $K$%
-matrix with the following elements: 
\begin{eqnarray*}
k_{12}(u) &=&\frac{1}{\beta _{13}}\Gamma _{1}(u)k_{13}(u),\quad k_{21}(u)=%
\frac{\beta _{21}}{\beta _{12}\beta _{13}}\Gamma _{1}(u)k_{13}(u), \\
k_{23}(u) &=&\frac{1}{\beta _{13}}\Gamma _{2}(u)k_{13}(u),\quad k_{32}(u)=%
\frac{\epsilon \beta _{21}}{\beta _{12}\beta _{13}}\Gamma _{2}(u)k_{13}(u),
\\
k_{31}(u) &=&\frac{\epsilon \beta _{21}^{2}}{\beta _{12}^{2}}k_{13}(u), \\
k_{13}(u) &=&\frac{1}{{\cal D}}[-2\beta _{12}^{2}\beta
_{13}x_{2}(x_{5}w(x_{2},y_{5})+y_{5}w(x_{2},x_{5}))],
\end{eqnarray*}
\begin{eqnarray}
k_{11}(u) &=&\frac{1}{{\cal D}}\left\{ (2w(x_{2},x_{5})+\beta
_{11}x_{2}x_{5})\left[ \beta _{13}\beta _{21}(\Gamma _{1}x_{6}y_{5}+\Gamma
_{2}x_{2}x_{7})\right. \right.  \nonumber \\
&&\left. -\epsilon \beta _{12}\beta _{13}\beta _{21}x_{1}x_{6}+2\beta
_{12}\Gamma _{1}w(x_{1},x_{2})\right] -\beta _{21}x_{2}x_{5}\left[ \epsilon
\beta _{13}^{2}\beta _{21}x_{2}x_{5}\right.  \nonumber \\
&&\left. \left. +\beta _{12}^{2}(\epsilon \Gamma _{1}x_{2}x_{4}-\Gamma
_{1}x_{1}x_{2}+\epsilon \Gamma _{2}x_{5}x_{6})-\epsilon \beta _{12}\beta
_{23}(\Gamma _{1}x_{6}y_{5}+\Gamma _{2}x_{2}x_{7})\right] \right\} . 
\nonumber \\
&&  \label{eq2.25}
\end{eqnarray}
where 
\begin{equation}
\Gamma _{1}(u)=\frac{\beta _{12}x_{3}x_{5}-\epsilon \beta _{23}x_{2}x_{6}}{%
x_{1}x_{3}-x_{3}^{2}+\epsilon x_{6}y_{6}}\quad {\rm and}\quad \Gamma _{2}(u)=%
\frac{\beta _{23}x_{3}y_{5}-\beta _{12}x_{2}y_{6}}{x_{1}x_{3}-x_{3}^{2}+%
\epsilon x_{6}y_{6}},  \label{eq2.19}
\end{equation}
and 
\begin{eqnarray}
{\cal D} &=&(2w(x_{2},y_{5})-\beta _{11}x_{2}y_{5})\left\{ \epsilon \beta
_{13}\beta _{21}[\Gamma _{1}x_{6}y_{5}+\Gamma _{2}x_{2}x_{7}]+\beta
_{12}[2\Gamma _{1}w(x_{1},x_{2})\right.  \nonumber \\
&&\left. -\epsilon \beta _{13}\beta _{21}x_{1}x_{6}]\right\} +\epsilon \beta
_{21}x_{2}y_{5}\left\{ \beta _{13}^{2}\beta _{21}x_{2}x_{5}-\beta _{12}\beta
_{23}[\Gamma _{1}x_{6}y_{5}+\Gamma _{2}x_{2}x_{7}]\right.  \nonumber \\
&&\left. +\beta _{12}^{2}[\Gamma _{1}x_{2}x_{4}-\epsilon \Gamma
_{2}x_{1}x_{2}+\Gamma _{2}x_{5}x_{6}]\right\} .  \label{eq2.27}
\end{eqnarray}
Here we have used an additional identity 
\begin{equation}
(x_{2}^{2}-x_{1}^{2})w(x_{1},x_{2})=x_{1}x_{5}w(x_{2},y_{5})+x_{2}y_{5}w(x_{1},y_{5}),
\label{eq2.20}
\end{equation}
where $w(f,g)$ is the Wronskian of two functions $f(u)$ and $g(u)$%
\begin{equation}
w(f,g)=\frac{df(u)}{du}g(u)-f(u)\frac{dg(u)}{du}  \label{eq2.21}
\end{equation}
In (\ref{eq2.25}), $k_{22}(u)=1$ and we did not write the amplitude $%
k_{33}(u)$ but it is obtained from the amplitude $k_{11}(u)$ using the
interchange rule (\ref{eq2.6}).

Before we consider the type-II solution we observe that the parameters $%
\beta _{ij}$ are linked by the conditions (\ref{eq2.16}) and (\ref{eq2.17})
which together with the relation (\ref{eq2.15}) and the normalization (\ref
{eq2.3}) imply that 
\begin{equation}
{\rm if\quad }\beta _{ij}=0\quad {\rm than}\quad \left\{ 
\begin{array}{c}
k_{ij}(u)=0,\quad {\rm for}\quad i\neq j \\ 
{\rm or} \\ 
k_{ij}(u)=1,\quad {\rm for}\quad i=j
\end{array}
\right. .  \label{eq2.28}
\end{equation}
It means that vanish free parameters in a general solution give us $K$%
-matrices with different forms defined by their non-vanishing entries. These
reduced solutions will be presented in the appendices according to the
values of $\beta _{13}$ and $\beta _{31}$, {\it i.e.,} they are
sub-classified according to a of the four possibilities: (i) $\beta
_{13}\neq 0$ and $\beta _{31}\neq 0$; (ii) $\beta _{13}\neq 0$ and $\beta
_{31}=0$; (iii) $\beta _{13}=0$ and $\beta _{31}\neq 0$ and (iv) $\beta
_{13}=0$ and $\beta _{31}=0$.

\subsection{Type-II solution}

For the type-II solution we have $\beta _{12}=\beta _{21}=$ $\beta
_{23}=\beta _{32}=0$. It means that the $K$-matrix has the form 
\begin{equation}
K_{II}=\left( 
\begin{array}{ccc}
k_{11} & 0 & k_{13} \\ 
0 & 1 & 0 \\ 
k_{31} & 0 & k_{33}
\end{array}
\right) .  \label{eq2.29}
\end{equation}
Now the boundary {\small YB} equation (\ref{int.7}) is composite of $28$
reflection equations collected in eight blocks $B[1,3]$, $B[1,5]$, $B[1,7]$, 
$B[2,4]$, $B[2,6]$, $B[3,5]$, $B[3,7]$ and $B[4,6]$.

From the blocks 
\begin{equation}
B[2,4]=[\beta _{11}x_{2}y_{5}-2w(x_{2},y_{5})]k_{11}+\beta
_{31}x_{2}x_{5}k_{13}+[\beta _{11}x_{2}x_{5}+2w(x_{2},x_{5})],
\label{eq2.30}
\end{equation}
and 
\begin{equation}
B[2,6]=\beta _{13}x_{2}y_{5}k_{11}+[\beta
_{33}x_{2}x_{5}-2w(x_{2},x_{5})]k_{13}+\beta _{13}x_{2}x_{5},  \label{eq2.31}
\end{equation}
we get the following four-parameter solution 
\begin{eqnarray}
k_{11} &=&-\frac{[\beta _{11}x_{2}x_{5}+2w(x_{2},x_{5})][\beta
_{33}x_{2}x_{5}-2w(x_{2},x_{5})]-\beta _{13}\beta _{31}x_{2}^{2}x_{5}^{2}}{%
[\beta _{11}x_{2}y_{5}-2w(x_{2},y_{5})][\beta
_{33}x_{2}x_{5}-2w(x_{2},x_{5})]-\beta _{13}\beta _{31}x_{2}^{2}x_{5}y_{5}},
\nonumber \\
k_{33} &=&-\frac{[\beta _{11}x_{2}y_{5}+2w(x_{2},y_{5})][\beta
_{33}x_{2}y_{5}-2w(x_{2},y_{5})]-\beta _{13}\beta _{31}x_{2}^{2}y_{5}^{2}}{%
[\beta _{11}x_{2}y_{5}-2w(x_{2},y_{5})][\beta
_{33}x_{2}x_{5}-2w(x_{2},x_{5})]-\beta _{13}\beta _{31}x_{2}^{2}x_{5}y_{5}},
\nonumber \\
k_{13} &=&\frac{2\beta
_{13}[x_{2}x_{5}w(x_{2},y_{5})+x_{2}y_{5}w(x_{2},x_{5})]}{[\beta
_{11}x_{2}y_{5}-2w(x_{2},y_{5})][\beta
_{33}x_{2}x_{5}-2w(x_{2},x_{5})]-\beta _{13}\beta _{31}x_{2}^{2}x_{5}y_{5}},
\nonumber \\
k_{31} &=&\frac{\beta _{31}}{\beta _{13}}k_{13}.  \label{eq2.32}
\end{eqnarray}
Substituting (\ref{eq2.32}) into the remaining blocks we will get constraint
equations involving $\beta _{11},\beta _{33},\beta _{13}$ and $\beta _{31}$.

Having now built \ a common ground for all elements of the reflection
matrix, we may proceed \ to find explicitly all regular $K$-matrices for
each model.

We begin with the {\small ZF} model because of its simplicity and because
all $K$-matrices are known from reference \cite{Inami}{\it . }Doing this,
besides confirming our result, we can compare its degree of simplicity.

\section{Regular K-matrix for the ZF model}

The simplest $19$-vertex model is the {\small ZF} model. This vertex model
is defined in terms of the Boltzmann weights given by the $R$-matrix of spin-%
$1$ representation of $U_{q}(\stackrel{\frown }{sl_{2}})$, or $A_{1}^{(1)}$
model. It is the trigonometric solution of the {\small YB} equation (\ref
{int.1}). $R(u)$ has the form (\ref{eq2.1}) with non-zero entries \cite{ZF}\ 
\begin{eqnarray}
x_{1}(u) &=&\sinh (u+\eta )\sinh (u+2\eta ),\quad x_{2}(u)=\sinh u\sinh
(u+\eta ),  \nonumber \\
x_{3}(u) &=&\sinh u\sinh (u-\eta ),\quad x_{5}(u)=y_{5}(u)=\sinh (u+\eta
)\sinh 2\eta ,  \nonumber \\
x_{6}(u) &=&y_{6}(u)=\sinh u\sinh 2\eta ,\quad x_{7}(u)=y_{7}(u)=\sinh \eta
\sinh 2\eta ,  \nonumber \\
x_{4}(u) &=&x_{2}(u)+x_{7}(u).  \label{eq3.1}
\end{eqnarray}
In this model, the Wronskians (\ref{eq2.21}) used in (\ref{eq2.25}) and (\ref
{eq2.32}) are given by 
\begin{eqnarray}
w(x_{1},x_{2}) &=&-\sinh 2\eta \sinh ^{2}(u+\eta ),  \nonumber \\
w(x_{2},x_{5}) &=&w(x_{2},y_{5})=\cosh u\sinh 2\eta \sinh ^{2}(u+\eta ),
\label{eq3.2}
\end{eqnarray}
and the $\Gamma $-functions (\ref{eq2.19}) have the form 
\begin{equation}
\Gamma _{1}(u)=\frac{\beta _{12}\sinh (u-\eta )-\beta _{23}\sinh u}{\sinh
(2u-\eta )},\quad \Gamma _{2}(u)=\frac{\beta _{23}\sinh (u-\eta )-\beta
_{12}\sinh u}{\sinh (2u-\eta )}.  \label{eq3.3}
\end{equation}
Substituting these expressions into the constraint equation (\ref{eq2.22})
we find in the first step that 
\begin{equation}
\sinh 2\eta \sinh ^{2}u\sinh ^{3}(u+\eta )F(u,\eta )=0,  \label{eq3.4}
\end{equation}
where 
\begin{eqnarray}
F(u,\eta ) &=&-2\beta _{12}^{2}\beta _{23}^{2}\sinh \eta \sinh u\cosh
(u-\eta )+\beta _{13}^{2}\beta _{21}\beta _{23}\sinh \eta \sinh 2\eta  
\nonumber \\
&&-2\beta _{12}^{2}\beta _{13}\sinh 2\eta \cosh (2u-\eta )+\beta
_{12}^{2}\beta _{13}\beta _{33}\sinh 2\eta \sinh (2u-\eta )  \nonumber \\
&&-2\beta _{12}\beta _{13}\beta _{23}\sinh 2\eta +\beta _{12}\beta
_{13}^{2}\beta _{21}\sinh 2\eta \sinh 2u  \nonumber \\
&&+\beta _{12}^{3}\beta _{23}\sinh \eta \sinh (2u-2\eta ).  \label{eq3.5}
\end{eqnarray}
This equation should hold for all values of $u$. In particular, for $u=0$ we
must have $F(0,\eta )=0$ from which we find $\beta _{21}$. Substituting $%
\beta _{21}$ into (\ref{eq3.5}) we can still write $F(u,\eta )=\sinh u\
G(u,\eta )=0$. From the condition $G(0,\eta )=0$ we find $\beta _{33}$.
Combining these results with the relations (\ref{eq2.16}) and (\ref{eq2.17})
we have

\begin{eqnarray}
\beta _{21} &=&\frac{\beta _{12}^{2}\beta _{23}}{\beta _{13}^{2}}\frac{1}{%
2\cosh \eta }+\frac{\beta _{12}}{\beta _{13}}\frac{2}{\sinh \eta }, 
\nonumber \\
\beta _{33} &=&\frac{\beta _{12}^{2}}{\beta _{13}}\frac{1}{2\cosh \eta }-2%
\frac{\cosh \eta }{\sinh \eta }-\frac{\beta _{12}\beta _{23}}{\beta _{13}}, 
\nonumber \\
\beta _{11} &=&\frac{\beta _{23}^{2}}{\beta _{13}}\frac{1}{2\cosh \eta }-2%
\frac{\cosh \eta }{\sinh \eta }-\frac{\beta _{12}\beta _{23}}{\beta _{13}}, 
\nonumber \\
\beta _{31} &=&\frac{\beta _{21}^{2}}{\beta _{12}^{2}}\beta _{13},\qquad
\beta _{32}=\frac{\beta _{21}\beta _{23}}{\beta _{12}}.  \label{eq3.6}
\end{eqnarray}
The parameter $\beta _{11}$ is fixed by the interchange rule (\ref{eq2.6})
or by direct computation with the constraint equation (\ref{eq2.23}). These
three free parameter expressions solve all constraint equations which come
from the expressions for $k_{13}(u)$ discussed \ in the previous section.

Substituting (\ref{eq3.6}) into the relations for type-I solution (\ref
{eq2.25}) we find the complete reflection $K$-matrix for the {\small ZF}
model:

\begin{eqnarray*}
k_{12}(u) &=&\frac{1}{\beta _{13}}\Gamma _{1}(u)k_{13}(u),\quad k_{21}(u)=%
\frac{\beta _{21}}{\beta _{12}\beta _{13}}\Gamma _{1}(u)k_{13}(u), \\
k_{23}(u) &=&\frac{1}{\beta _{13}}\Gamma _{2}(u)k_{13}(u),\quad k_{32}(u)=%
\frac{\beta _{21}}{\beta _{12}\beta _{13}}\Gamma _{2}(u)k_{13}(u), \\
k_{31}(u) &=&\frac{\beta _{21}^{2}}{\beta _{12}^{2}}k_{13}(u), \\
k_{13}(u) &=&\frac{\beta _{13}^{2}\sinh 2\eta \sinh 2u\sinh (2u-\eta )}{%
{\cal D}(u,\eta )},
\end{eqnarray*}
\begin{eqnarray}
k_{11}(u) &=&\frac{1}{{\cal D}(u,\eta )}\left\{ \sinh 2\eta (\beta
_{12}\beta _{23}\sinh u\sinh (u-\eta )-2\beta _{13}\sinh (2u-\eta ))\right. 
\nonumber \\
&&\left. -\sinh \eta \sinh u(\beta _{12}^{2}\sinh (u-\eta )+\beta
_{23}^{2}\sinh (u+\eta ))\right\} ,  \nonumber \\
&&  \nonumber \\
k_{33}(u) &=&\frac{1}{{\cal D}(u,\eta )}\left\{ \sinh 2\eta (\beta
_{12}\beta _{23}\sinh u\sinh (u-\eta )-2\beta _{13}\sinh (2u-\eta ))\right. 
\nonumber \\
&&-\left. \sinh \eta \sinh u(\beta _{12}^{2}\sinh (u+\eta )+\beta
_{23}^{2}\sinh (u-\eta ))\right\} ,  \label{eq3.7}
\end{eqnarray}
where

\begin{eqnarray}
{\cal D}(u,\eta ) &=&\beta _{12}\beta _{23}(\sinh \eta \sinh ^{2}2u+\sinh
u\sinh (u-\eta )\sinh 2\eta )  \nonumber \\
&&-(\beta _{12}^{2}+\beta _{23}^{2})\sinh u\sinh (u-\eta )\sinh \eta 
\nonumber \\
&&+4\beta _{13}\cosh \eta \sinh (2u-\eta )\sinh (2u+\eta ).  \label{eq3.8}
\end{eqnarray}
It is a regular solution of the {\small RE} equation (\ref{int.7}) with
three free parameters $\beta _{12},\beta _{13}$ and $\beta _{32}$.

Using the relations 
\begin{eqnarray}
\beta _{12} &=&-\frac{4\mu \stackrel{\sim }{\mu }\sinh (\frac{\eta }{2}%
-\zeta )\cosh \eta }{2\sinh (\frac{\eta }{2}+\zeta )\sinh (\frac{\eta }{2}%
-\zeta )\cosh \eta +\mu \stackrel{\sim }{\mu }\sinh ^{2}\eta },  \nonumber \\
\beta _{13} &=&\frac{2\mu ^{2}\sinh \eta }{2\sinh (\frac{\eta }{2}+\zeta
)\sinh (\frac{\eta }{2}-\zeta )\cosh \eta +\mu \stackrel{\sim }{\mu }\sinh
^{2}\eta },  \nonumber \\
\beta _{32} &=&\frac{4\mu \stackrel{\sim }{\mu }\sinh (\frac{\eta }{2}+\zeta
)\cosh \eta }{2\sinh (\frac{\eta }{2}+\zeta )\sinh (\frac{\eta }{2}-\zeta
)\cosh \eta +\mu \stackrel{\sim }{\mu }\sinh ^{2}\eta },  \label{eq3.9}
\end{eqnarray}
where\ $\mu ,\stackrel{\sim }{\mu }$ and $\zeta $ are new free parameters,
we will reproduce the solution derived by Inami {\it et al} \cite{Inami}
using a different approach to solve the {\small RE.}

Now we proceed to see the role of the type-II solution (\ref{eq2.28}) in the 
{\small ZF} model. Taking into account the data of this model we substitute (%
\ref{eq2.28}) into the equations $E[1,5]=0$ and $E[5,9]=0$ of the block $%
B[1,5]$ and write them in factored form 
\begin{equation}
2\sinh ^{2}u\cosh u\sinh ^{4}(u+\eta )F_{i}(u,\eta )=0,\qquad i=1,2
\label{eq3.11}
\end{equation}
where 
\begin{eqnarray}
F_{1}(u,\eta ) &=&2\beta _{11}(\sinh (u+3\eta )-\cosh u\sinh \eta )-2\beta
_{33}\sinh u\cosh \eta  \nonumber \\
&&+(\beta _{11}\beta _{33}-\beta _{13}\beta _{31})\sinh u\sinh \eta +4\cosh
(u+3\eta )  \nonumber \\
&&+4\cosh u\cosh \eta ,  \label{eq3.12}
\end{eqnarray}
and $F_{2}(u,\eta )$ is obtained from $F_{1}(u,\eta )$ by interchanging the
indices $1\leftrightarrow 3$. The conditions $F_{1}(0,\eta )=F_{2}(0,\eta
)=0 $ imply that 
\begin{equation}
\beta _{11}=\beta _{33}=-2\frac{\cosh \eta }{\sinh \eta }.  \label{eq3.13}
\end{equation}
If now we recall the relations (\ref{eq3.6}) we will see that this is not a
new solution but the one-parameter solution obtained by reduction of the
type-I solution through the limit $\beta _{12}\rightarrow 0$ and $\beta
_{23}\rightarrow 0$. The corresponding $K$ matrix has the form (\ref{eq2.29}%
) with non-zero entries 
\begin{eqnarray}
k_{13}(u) &=&\frac{1}{2}\frac{\beta _{13}\sinh \eta \sinh 2u}{\sinh (2u+\eta
)},\quad k_{31}(u)=\frac{1}{2}\frac{\beta _{31}\sinh \eta \sinh 2u}{\sinh
(2u+\eta )}  \nonumber \\
k_{11}(u) &=&k_{33}(u)=\frac{\sinh \eta }{\sinh (2u-\eta )},  \nonumber \\
\beta _{31} &=&\frac{4}{\beta _{13}\sinh ^{2}\eta }.  \label{eq3.14}
\end{eqnarray}
The relation between $\beta _{31}$ and $\beta _{13}$ in (\ref{eq3.14}) was
obtained using the equation $E[2,4]=0$.

In the appendix A we will present the remaining reduced solutions which can
be obtained form the complete reflection $K$-matrix of the {\small ZF} model.

\section{Regular K-matrices for the IK model}

The solution of the {\small YB} equation corresponding to $A_{2}^{(2)}$ in
the fundamental representation was found by Izergin and Korepin \cite{IK}.
The $R$-matrix has the form (\ref{eq2.1}) with non-zero entries 
\begin{eqnarray}
x_{1}(u) &=&2\sinh (\frac{u}{2}-2\eta )\cosh (\frac{u}{2}-3\eta ),\quad
x_{2}(u)=2\sinh \frac{u}{2}\cosh (\frac{u}{2}-3\eta ),  \nonumber \\
x_{3}(u) &=&2\sinh \frac{u}{2}\cosh (\frac{u}{2}-\eta ),  \nonumber \\
x_{4}(u) &=&2\sinh \frac{u}{2}\cosh (\frac{u}{2}-3\eta )-2\sinh 2\eta \cosh
3\eta ,  \nonumber \\
x_{5}(u) &=&2{\rm e}^{-\frac{1}{2}u}\sinh 2\eta \cosh (\frac{u}{2}-3\eta
),\quad y_{5}(u)=-{\rm e}^{u}x_{5}(u),  \nonumber \\
x_{6}(u) &=&2{\rm e}^{-\frac{1}{2}u+2\eta }\sinh 2\eta \sinh \frac{u}{2}%
,\quad \qquad y_{6}(u)={\rm e}^{u-4\eta }x_{6}(u),  \nonumber \\
x_{7}(u) &=&-2{\rm e}^{-\frac{1}{2}u}\sinh 2\eta \lbrack \cosh (\frac{u}{2}%
-3\eta )+{\rm e}^{\eta }\sinh \frac{u}{2}],  \nonumber \\
y_{7}(u) &=&-2{\rm e}^{\frac{1}{2}u}\sinh 2\eta \lbrack \cosh (\frac{u}{2}%
-3\eta )-{\rm e}^{-\eta }\sinh \frac{u}{2}].  \label{eq5.1}
\end{eqnarray}
For this model the Wronskians are given by 
\begin{equation}
w(x_{1},x_{2})=2\sinh 2\eta \cosh ^{2}(\frac{u}{2}-3\eta ),\
w(x_{2},x_{5})=w(x_{2},y_{5})=-w(x_{1},x_{2}).  \label{eq5.2}
\end{equation}

A partial classification already exists for the reflection $K$-matrices of
the {\small IK} model. Here we will use the procedure discussed in the
section $2$ to extend the results presented in \cite{Kim}.

Following the steps used in the construction of the solution for the {\small %
ZF} model, we start by considering the constraint equations (\ref{eq2.22})
and (\ref{eq2.23}) with the data of the {\small IK} model. After we find the
expressions to $\beta _{21}$ and $\beta _{33}$ through the factorization
procedure described in the section $2$, we will arrive at the following
equation 
\begin{equation}
(\beta _{12}^{2}+{\rm e}^{4\eta }\beta _{23}^{2})\cosh 2\eta \sinh \frac{u}{2%
}(\beta _{11}{\rm e}^{\frac{1}{2}u}\sinh \frac{u}{2}-1)=0,  \label{eq5.3}
\end{equation}
which has an interchanged partner by the rule (\ref{eq2.7}). Both equations
must be valid for all values of $u$.

For $\beta _{12},\beta _{23},\beta _{21}$ and $\beta _{32}$ different of
zero we have two solutions: 
\begin{equation}
\{\beta _{23}=i{\rm e}^{-2\eta }\beta _{12},\ \beta _{32}=i{\rm e}^{-2\eta
}\beta _{21}\}\quad {\rm and}\quad \{\beta _{23}=-i{\rm e}^{-2\eta }\beta
_{12},\ \beta _{32}=-i{\rm e}^{-2\eta }\beta _{21}\}.  \label{eq5.3a}
\end{equation}
The third solution is $\beta _{12}=\beta _{23}=\beta _{21}=\beta _{32}=0$.

\subsection{Type-I Solutions}

Due to (\ref{eq5.3a}), the constraint equations (\ref{eq2.22}) and (\ref
{eq2.23}) for the {\small IK} model have two type-I solutions. Their
parameters are given by 
\begin{eqnarray}
\beta _{11} &=&\mp i\frac{\beta _{12}^{2}}{\beta _{13}}\frac{2{\rm e}%
^{-2\eta }\cosh \eta \pm i}{2\cosh \eta }\mp i{\rm e}^{-\eta }\frac{1\mp
i\sinh \eta }{(\cosh \eta \mp i\sinh 2\eta )\cosh \eta },  \nonumber \\
\beta _{33} &=&\mp i\frac{\beta _{12}^{2}}{\beta _{13}}\frac{2{\rm e}%
^{-4\eta }\cosh \eta \mp i}{2\cosh \eta }\mp i{\rm e}^{\eta }\frac{1\mp
i\sinh \eta }{(\cosh \eta \mp i\sinh 2\eta )\cosh \eta },  \nonumber \\
\beta _{21} &=&\pm i\frac{\beta _{12}^{3}}{\beta _{13}^{2}}\frac{{\rm e}%
^{-2\eta }}{2\cosh \eta }\pm i\frac{\beta _{12}}{\beta _{13}}\frac{1\mp
i\sinh \eta }{(\cosh \eta \mp i\sinh 2\eta )\cosh \eta },  \nonumber \\
\beta _{23} &=&\pm i\beta _{12}{\rm e}^{-2\eta },\quad \beta _{32}=\pm
i\beta _{21}{\rm e}^{-2\eta },\quad \beta _{31}=\frac{\beta _{21}^{2}}{\beta
_{12}^{2}}\beta _{13},  \label{eq5.4}
\end{eqnarray}
and their $\Gamma $-functions are given by 
\begin{equation}
\Gamma _{1}(u)=\beta _{12}{\rm e}^{-\frac{1}{2}u}\frac{\cosh (\frac{u}{2}%
-\eta )\pm i\sinh (\frac{u}{2})}{\cosh (u-\eta )},\quad \Gamma _{2}(u)=\pm i%
{\rm e}^{u-2\eta }\Gamma _{1}(u).  \label{eq5.5}
\end{equation}
It means that these solutions have two free parameters, $\beta _{12}$ and $%
\beta _{13}$.

To write both type-I solutions of the {\small IK} model is a little bit
cumbersome. Therefore, in order to save notation, we will only write one
solution and take the conjugate of it to get the second solution.

For the first solution we find that the off-diagonal components of the $K$%
-matrix are given by

\begin{eqnarray}
k_{12}(u) &=&\frac{1}{\beta _{13}}\Gamma _{1}(u)k_{13}(u),\quad k_{21}(u)=%
\frac{\beta _{21}}{\beta _{12}\beta _{13}}\Gamma _{1}(u)k_{13}(u),  \nonumber
\\
k_{23}(u) &=&\frac{1}{\beta _{13}}\Gamma _{2}(u)k_{13}(u),\quad k_{32}(u)=%
\frac{\beta _{21}}{\beta _{12}\beta _{13}}\Gamma _{2}(u)k_{13}(u),  \nonumber
\\
k_{31}(u) &=&\frac{\beta _{21}^{2}}{\beta _{12}^{2}}k_{13}(u),  \nonumber \\
k_{13}(u) &=&\frac{\beta _{13}\sinh u\cosh 3\eta \lbrack \cosh (\frac{u}{2}%
-\eta )-i\sinh \frac{u}{2}]}{\cosh \eta \lbrack \cosh (\frac{u}{2}+3\eta
)-i\sinh \frac{u}{2}]+\cosh 3\eta \sinh \frac{u}{2}{\cal F}(u,\eta )},
\label{eq5.6}
\end{eqnarray}
and the diagonal components are given by 
\begin{eqnarray}
k_{11}(u) &=&\frac{{\rm e}^{-u}\cosh \eta \lbrack \cosh (\frac{u}{2}-3\eta
)+i\sinh \frac{u}{2}]+{\rm e}^{-\frac{1}{2}u}\cosh 3\eta \sinh \frac{u}{2}%
\Omega _{-}(u,\eta )}{\cosh \eta \lbrack \cosh (\frac{u}{2}+3\eta )-i\sinh 
\frac{u}{2}]+\cosh 3\eta \sinh \frac{u}{2}{\cal F}(u,\eta )},  \nonumber \\
k_{33}(u) &=&\frac{{\rm e}^{u}\cosh \eta \lbrack \cosh (\frac{u}{2}-3\eta
)+i\sinh \frac{u}{2}]-{\rm e}^{\frac{1}{2}u}\cosh 3\eta \sinh \frac{u}{2}%
\Omega _{+}(u,\eta )}{\cosh \eta \lbrack \cosh (\frac{u}{2}+3\eta )-i\sinh 
\frac{u}{2}]+\cosh 3\eta \sinh \frac{u}{2}{\cal F}(u,\eta )}.  \label{eq5.7}
\end{eqnarray}
Here $\Omega _{\mp }(u,\eta )$ and ${\cal F}(u,\eta )$ are two-parameter
functions defined by 
\begin{eqnarray}
\Omega _{\mp }(u,\eta ) &=&[\cosh (2\eta )\cosh (\frac{u}{2}-\eta )\mp \frac{%
i}{2}(e^{\mp \frac{1}{2}u}+e^{\pm \frac{1}{2}u}\cosh 2\eta )]f(\eta ), 
\nonumber \\
{\cal F}(u,\eta ) &=&[\sinh (\frac{u}{2}-2\eta )\cosh (\frac{u}{2}+\eta
)+i\cosh ^{2}(\frac{u}{2}-\eta )]f(\eta ),  \label{eq5.8}
\end{eqnarray}
where 
\begin{equation}
f(\eta )=\frac{\beta _{12}^{2}}{\beta _{13}}\frac{{\rm e}^{-2\eta }}{\cosh
\eta }+\frac{2}{\cosh 2\eta -i\sinh \eta }.  \label{eq5.9}
\end{equation}
The second type-I solution can be obtained from these expressions by
interchanging $\pm i\leftrightarrow \mp i$. Of course there are many ways to
write down these expressions. However, we chose to write them in this form
in order to get more easily their reduced solutions in the appendix B.

\subsection{Type-II Solution}

The procedure to obtain type-I and type-II solutions are different by
construction. To get the type-II we recall the expressions (\ref{eq2.32})
with the data of the {\small IK} model. The result is a two-parameter $K$%
-matrix which has the form (\ref{eq2.29}) with non-zero entries 
\begin{eqnarray}
k_{31}(u) &=&\frac{\beta _{31}}{\beta _{13}}k_{13}(u),  \nonumber \\
k_{13}(u) &=&\frac{\beta _{13}\sinh u}{1-2\beta _{11}{\rm e}^{\eta }\sinh 
\frac{u}{2}\cosh (\frac{u}{2}-\eta )},  \nonumber \\
k_{11}(u) &=&\frac{1-2\beta _{11}{\rm e}^{-\frac{1}{2}u+\eta }\sinh \frac{u}{%
2}\sinh \eta }{1-2\beta _{11}{\rm e}^{\eta }\sinh \frac{u}{2}\cosh (\frac{u}{%
2}-\eta )},  \nonumber \\
k_{33}(u) &=&\frac{1+2\beta _{11}{\rm e}^{\frac{1}{2}u+\eta }\sinh \frac{u}{2%
}\sinh \eta }{1-2\beta _{11}{\rm e}^{\eta }\sinh \frac{u}{2}\cosh (\frac{u}{2%
}-\eta )}.  \label{eq5.19}
\end{eqnarray}
Substituting (\ref{eq5.19}) into the equation $E[1,5]=0$ we will find the
following relations for the parameters 
\begin{equation}
\beta _{33}={\rm e}^{2\eta }\beta _{11}\quad {\rm and}\quad \beta _{31}=%
\frac{1}{\beta _{13}}{\rm e}^{2\eta }\beta _{11}^{2}  \label{eq5.20}
\end{equation}
Here we observe that this solution was already obtained in \cite{Kim}.

To summarize, we have obtained three regular general solutions with two free
parameters of the {\small RE} for the {\small IK} model. The two first are
type-I solution and the last is a type-II solution. If we compare the number
of type-I solutions of the {\small IK} model with the number found for the 
{\small ZF} model we can see that the reduction of the number of free
parameters from three for two increase the number of possible solutions as
much as the possible manners of find the last parameter.

\section{Regular K-matrix for the $sl(2|1)$ model}

The solution of the graded {\small YB} equation corresponding to $sl(2|1)$
in the fundamental representation has the form (\ref{eq2.1}) with non-zero
entries \cite{Kulish1,Bazhanov2,Kulish4}: 
\begin{eqnarray}
x_{1}(u) &=&\sinh (u+2\eta )\cosh (u+\eta ),\quad x_{2}(u)=\sinh u\cosh
(u+\eta ),  \nonumber \\
x_{3}(u) &=&\sinh u\cosh (u-\eta ),\quad x_{5}(u)=y_{5}(u)=\sinh 2\eta \cosh
(u+\eta ),  \nonumber \\
x_{6}(u) &=&y_{6}(u)=\sinh u\sinh 2\eta ,\quad x_{7}(u)=y_{7}(u)=\sinh 2\eta
\cosh \eta ,  \nonumber \\
x_{4}(u) &=&x_{2}(u)-x_{7}(u).  \label{eq4.1}
\end{eqnarray}
For this model the Wronskians and the $\Gamma $-functions are given by 
\[
w(x_{1},x_{2})=-\sinh 2\eta \cosh ^{2}(u+\eta ),\ w(x_{2},x_{5})=\sinh 2\eta
\cosh u\cosh ^{2}(u+\eta ), 
\]
\begin{equation}
\Gamma _{1}(u)=\frac{\beta _{12}\cosh (u-\eta )+\beta _{23}\sinh u}{\cosh
(2u-\eta )},\ \Gamma _{2}(u)=\frac{\beta _{23}\cosh (u-\eta )-\beta
_{12}\sinh u}{\cosh (2u-\eta )}.  \label{eq4.2}
\end{equation}
From the similarity of these data with those of the {\small ZF }model, one
can expect that the steps which we proceeded to solve the $sl(2|1)$-model
are the same ones that we used in the {\small ZF} model. Therefore we need
only to present the solution.

The parameters are given by 
\begin{eqnarray}
\beta _{11} &=&-\frac{\beta _{23}^{2}}{\beta _{13}}\frac{1}{2\sinh \eta }-2%
\frac{\sinh \eta }{\cosh \eta }-\frac{\beta _{12}\beta _{23}}{\beta _{13}}, 
\nonumber \\
\beta _{33} &=&\frac{\beta _{12}^{2}}{\beta _{13}}\frac{1}{2\sinh \eta }-2%
\frac{\sinh \eta }{\cosh \eta }-\frac{\beta _{12}\beta _{23}}{\beta _{13}}, 
\nonumber \\
\beta _{21} &=&-\frac{\beta _{12}^{2}\beta _{23}}{\beta _{13}^{2}}\frac{1}{%
2\sinh \eta }-\frac{\beta _{12}}{\beta _{13}}\frac{2}{\cosh \eta }, 
\nonumber \\
\beta _{31} &=&-\frac{\beta _{21}^{2}}{\beta _{12}^{2}}\beta _{13},\qquad
\beta _{32}=-\frac{\beta _{21}\beta _{23}}{\beta _{12}},  \label{eq4.3}
\end{eqnarray}
which results in a type-I solution with three free parameters. The
off-diagonal elements are given by 
\begin{eqnarray}
k_{12}(u) &=&\frac{1}{\beta _{13}}\Gamma _{1}(u)k_{13}(u),\quad k_{21}(u)=%
\frac{\beta _{21}}{\beta _{12}\beta _{13}}\Gamma _{1}(u)k_{13}(u),  \nonumber
\\
k_{23}(u) &=&\frac{1}{\beta _{13}}\Gamma _{2}(u)k_{13}(u),\quad k_{32}(u)=-%
\frac{\beta _{21}}{\beta _{12}\beta _{13}}\Gamma _{2}(u)k_{13}(u),  \nonumber
\\
k_{31}(u) &=&-\frac{\beta _{21}^{2}}{\beta _{12}^{2}}k_{13}(u),  \nonumber \\
k_{13}(u) &=&\frac{\beta _{13}^{2}\sinh 2\eta \sinh 2u\cosh (2u-\eta )}{%
{\cal D}(u,\eta )},  \label{eq4.4}
\end{eqnarray}
and the diagonal elements are given by 
\begin{eqnarray}
k_{11}(u) &=&\frac{1}{{\cal D}(u,\eta )}\left\{ \sinh 2\eta (-\beta
_{12}\beta _{23}\sinh u\cosh (u-\eta )+2\beta _{13}\cosh (2u-\eta ))\right. 
\nonumber \\
&&\left. -\cosh \eta \sinh u(\beta _{12}^{2}\sinh (u-\eta )+\beta
_{23}^{2}\cosh (u+\eta ))\right\} ,  \nonumber \\
&&  \nonumber \\
k_{33}(u) &=&\frac{1}{{\cal D}(u,\eta )}\left\{ \sinh 2\eta (-\beta
_{12}\beta _{23}\sinh u\cosh (u-\eta )+2\beta _{13}\cosh (2u-\eta ))\right. 
\nonumber \\
&&+\left. \cosh \eta \sinh u(\beta _{12}^{2}\cosh (u+\eta )+\beta
_{23}^{2}\cosh (u-\eta ))\right\} ,  \label{eq4.5}
\end{eqnarray}
where

\begin{eqnarray}
{\cal D}(u,\eta ) &=&2\beta _{12}\beta _{23}\sinh u\cosh \eta (\sinh u+\cosh
(2u-\eta )\sinh (u+\eta ))  \nonumber \\
&&+(\beta _{12}^{2}-\beta _{23}^{2})\sinh u\cosh (u-\eta )\cosh \eta 
\nonumber \\
&&+4\beta _{13}\sinh \eta \cosh (2u-\eta )\cosh (2u+\eta ).  \label{eq4.6}
\end{eqnarray}
Regular reduced solutions from this complete reflection $K$-matrix are
presented in the appendix C.

\section{Regular K-matrices for the $osp(2|1)$ model}

The trigonometric solution of the graded {\small YB} equation for the
fundamental representation of the $osp(2|1)$ algebra was found by Bazhanov
and Shadrikov in \cite{Bazhanov2}. It has the form (\ref{eq2.1}) with 
\begin{eqnarray}
x_{1}(u) &=&\sinh (u+2\eta )\sinh (u+3\eta ),\quad x_{2}(u)=\sinh u\sinh
(u+3\eta ),  \nonumber \\
x_{3}(u) &=&\sinh u\sinh (u+\eta ),  \nonumber \\
x_{4}(u) &=&\sinh u\sinh (u+3\eta )-\sinh 2\eta \sinh 3\eta ,  \nonumber \\
x_{5}(u) &=&{\rm e}^{-u/3}\sinh 2\eta \sinh (u+3\eta ),\quad y_{5}(u)={\rm e}%
^{u/3}\sinh 2\eta \sinh (u+3\eta ),  \nonumber \\
x_{6}(u) &=&-{\rm e}^{-u/3-2\eta }\sinh 2\eta \sinh u,\quad y_{6}(u)={\rm e}%
^{u/3+2\eta }\sinh 2\eta \sinh u,  \nonumber \\
x_{7}(u) &=&{\rm e}^{u/3}\sinh 2\eta \left( \sinh (u+3\eta )+{\rm e}^{-\eta
}\sinh u\right) ,  \nonumber \\
y_{7}(u) &=&{\rm e}^{-u/3}\sinh 2\eta \left( \sinh (u+3\eta )+{\rm e}^{\eta
}\sinh u\right) .  \label{eq6.1}
\end{eqnarray}
For this model the Wronskians are given by 
\begin{eqnarray}
w(x_{1},x_{2}) &=&-\sinh 2\eta \sinh ^{2}(u+3\eta ),  \nonumber \\
w(x_{2},x_{5}) &=&\frac{1}{3}(2{\rm e}^{\frac{2}{3}u}+{\rm e}^{-\frac{4}{3}%
u})\sinh 2\eta \sinh ^{2}(u+3\eta ),  \nonumber \\
w(x_{2},y_{5}) &=&\frac{1}{3}(2{\rm e}^{-\frac{2}{3}u}+{\rm e}^{\frac{4}{3}%
u})\sinh 2\eta \sinh ^{2}(u+3\eta ).  \label{eq6.2}
\end{eqnarray}
Now we will calculate the solutions of the {\small RE} for the $osp(2|1)$
model. The procedure is a lot similar to that used to solve the {\small IK}
model. The final constraint equations like (\ref{eq5.3}) are 
\begin{equation}
\beta _{12}^{2}-{\rm e}^{-4\eta }\beta _{23}^{2}=0\quad {\rm and}\quad \beta
_{21}^{2}-{\rm e}^{-4\eta }\beta _{32}^{2}=0.  \label{eq6.3}
\end{equation}
For $\beta _{12},\beta _{23},\beta _{21}$ and $\beta _{32}$ different of
zero we have two type-I solutions 
\begin{equation}
\{\beta _{23}={\rm e}^{2\eta }\beta _{12},\ \beta _{32}=-{\rm e}^{2\eta
}\beta _{21}\}\quad {\rm and}\quad \{\beta _{23}=-{\rm e}^{2\eta }\beta
_{12},\ \beta _{32}={\rm e}^{-2\eta }\beta _{21}\}.  \label{eq6.3a}
\end{equation}
The relative sign between $\beta _{23}$ and $\beta _{32}$ in both type-I
solutions comes from the grading used. The third solution is $\beta
_{12}=\beta _{23}=\beta _{21}=\beta _{32}=0$, the type-II solution.

\subsection{Type-I Solutions}

For the $osp(2|1)$ model we have to write separately both type-I solutions.
The first has the following relations among the parameters 
\begin{eqnarray}
\beta _{11} &=&-\frac{\beta _{12}^{2}}{\beta _{13}}\frac{1+2{\rm e}^{2\eta
}\sinh \eta }{2\sinh \eta }+\frac{3{\rm e}^{\eta }+4\sinh \frac{1}{2}\eta
\sinh \frac{3}{2}\eta }{3\sinh \frac{1}{2}\eta \sinh \frac{3}{2}\eta }, 
\nonumber \\
\beta _{33} &=&\frac{\beta _{23}^{2}}{\beta _{13}}\frac{1-2{\rm e}^{-2\eta
}\sinh \eta }{2\sinh \eta }-\frac{3{\rm e}^{-\eta }+4\sinh \frac{1}{2}\eta
\sinh \frac{3}{2}\eta }{3\sinh \frac{1}{2}\eta \sinh \frac{3}{2}\eta }, 
\nonumber \\
\beta _{21} &=&\frac{\beta _{12}^{3}}{\beta _{13}^{2}}\frac{{\rm e}^{2\eta }%
}{2\sinh \eta }+\frac{\beta _{12}}{\beta _{13}}\frac{1}{\sinh \frac{1}{2}%
\eta \sinh \frac{3}{2}\eta },  \nonumber \\
\beta _{23} &=&{\rm e}^{2\eta }\beta _{12},\quad \beta _{32}=-{\rm e}^{2\eta
}\beta _{21},\quad \beta _{31}=-\frac{\beta _{21}^{2}}{\beta _{12}^{2}}\beta
_{13}.  \label{eq6.4}
\end{eqnarray}
The corresponding $\Gamma $-functions are given by 
\begin{equation}
\Gamma _{1}(u)=\beta _{12}e^{-\frac{1}{3}u}\frac{\sinh \frac{1}{2}\eta }{%
\sinh \frac{3}{2}\eta }\qquad {\rm and}\qquad \Gamma _{2}(u)={\rm e}^{\frac{2%
}{3}u+2\eta }\Gamma _{1}(u).  \label{eq6.5}
\end{equation}

Substituting these data into (\ref{eq2.25}) we will have a complete
reflection $K$-matrix with two free parameters. The corresponding entries
are 
\begin{eqnarray}
k_{12}(u) &=&\frac{1}{\beta _{13}}\Gamma _{1}(u)k_{13}(u),\quad k_{21}(u)=%
\frac{\beta _{21}}{\beta _{12}\beta _{13}}\Gamma _{1}(u)k_{13}(u),  \nonumber
\\
k_{23}(u) &=&\frac{1}{\beta _{13}}\Gamma _{2}(u)k_{13}(u),\quad k_{21}(u)=-%
\frac{\beta _{21}}{\beta _{12}\beta _{13}}\Gamma _{2}(u)k_{13}(u),  \nonumber
\\
k_{13}(u) &=&-\frac{1}{D_{1}(u,\eta )}\beta _{13}^{2}\sinh \frac{1}{2}\eta
\sinh \eta \sinh \frac{3}{2}\eta \sinh (u+\frac{1}{2}\eta )\sinh 2u, 
\nonumber \\
k_{13}(u) &=&-\frac{\beta _{21}^{2}}{\beta _{12}^{2}}k_{13}(u),  \nonumber \\
k_{11}(u) &=&\frac{1}{{\cal D}_{1}(u,\eta )}\left\{ \frac{1}{2}{\rm e}^{-%
\frac{2}{3}u}\sinh (u+\frac{1}{2}\eta )({\rm e}^{2u}\cosh 2\eta -\cosh \eta )%
{\cal F}_{1}(u,\eta )\right.  \nonumber \\
&&-\left. \beta _{12}^{2}{\rm e}^{-\frac{2}{3}u+2\eta }\sinh ^{3}\frac{1}{2}%
\eta \sinh \frac{3}{2}\eta \sinh (u+\frac{3}{2}\eta )\right\} ,  \nonumber \\
k_{33}(u) &=&\frac{1}{{\cal D}_{1}(u,\eta )}\left\{ \frac{1}{2}{\rm e}^{%
\frac{2}{3}u}\sinh (u+\frac{1}{2}\eta )({\rm e}^{-2u}\cosh 2\eta -\cosh \eta
){\cal F}_{1}(u,\eta )\right.  \nonumber \\
&&-\left. \beta _{23}^{2}{\rm e}^{\frac{2}{3}u-2\eta }\sinh ^{3}\frac{1}{2}%
\eta \sinh \frac{3}{2}\eta \sinh (u+\frac{3}{2}\eta )\right\} ,
\label{eq6.6}
\end{eqnarray}
where 
\begin{eqnarray}
{\cal D}_{1}(u,\eta ) &=&\beta _{12}^{2}{\rm e}^{2\eta }\sinh ^{3}\frac{1}{2}%
\eta \sinh \frac{3}{2}\eta \sinh (u-\frac{3}{2}\eta )  \nonumber \\
&&-\sinh (u-\frac{1}{2}\eta )\sinh (u+\frac{1}{2}\eta )\sinh (u+\frac{3}{2}%
\eta ){\cal F}_{1}(u,\eta ),  \label{eq6.8}
\end{eqnarray}
and 
\begin{equation}
{\cal F}_{1}(u,\eta )=\beta _{12}^{2}{\rm e}^{2\eta }\sinh \frac{1}{2}\eta
\sinh \frac{3}{2}\eta -2\beta _{13}\sinh \eta .  \label{eq6.9}
\end{equation}
For the second solution the parameters are given by

\begin{eqnarray}
\beta _{11} &=&-\frac{\beta _{12}^{2}}{\beta _{13}}\frac{1-2{\rm e}^{2\eta
}\sinh \eta }{2\sinh \eta }+\frac{3{\rm e}^{\eta }-4\cosh \frac{1}{2}\eta
\cosh \frac{3}{2}\eta }{3\cosh \frac{1}{2}\eta \cosh \frac{3}{2}\eta }, 
\nonumber \\
\beta _{33} &=&\frac{\beta _{23}^{2}}{\beta _{13}}\frac{1+2{\rm e}^{-2\eta
}\sinh \eta }{2\sinh \eta }+\frac{3{\rm e}^{-\eta }-4\cosh \frac{1}{2}\eta
\cosh \frac{3}{2}\eta }{3\cosh \frac{1}{2}\eta \cosh \frac{3}{2}\eta }, 
\nonumber \\
\beta _{21} &=&\frac{\beta _{12}^{3}}{\beta _{13}^{2}}\frac{{\rm e}^{2\eta }%
}{2\sinh \eta }-\frac{\beta _{12}}{\beta _{13}}\frac{1}{\cosh \frac{1}{2}%
\eta \cosh \frac{3}{2}\eta },  \nonumber \\
\beta _{23} &=&-{\rm e}^{2\eta }\beta _{12},\quad \beta _{32}={\rm e}^{2\eta
}\beta _{21},\quad \beta _{31}=-\frac{\beta _{21}^{2}}{\beta _{12}^{2}}\beta
_{13},  \label{eq6.10}
\end{eqnarray}
and we have new $\Gamma $-functions defined by 
\begin{equation}
\Gamma _{1}(u)=\beta _{12}e^{-\frac{1}{3}u}\frac{\cosh \frac{1}{2}\eta }{%
\cosh \frac{3}{2}\eta }\qquad {\rm and}\qquad \Gamma _{2}(u)=-{\rm e}^{\frac{%
2}{3}u+2\eta }\Gamma _{1}(u).  \label{eq6.11}
\end{equation}
This results into a second complete reflection $K$-matrix with entries 
\begin{eqnarray}
k_{12}(u) &=&\frac{1}{\beta _{13}}\Gamma _{1}(u)k_{13}(u),\quad k_{21}(u)=%
\frac{\beta _{21}}{\beta _{12}\beta _{13}}\Gamma _{1}(u)k_{13}(u),  \nonumber
\\
k_{23}(u) &=&\frac{1}{\beta _{13}}\Gamma _{2}(u)k_{13}(u),\quad k_{21}(u)=-%
\frac{\beta _{21}}{\beta _{12}\beta _{13}}\Gamma _{2}(u)k_{13}(u),  \nonumber
\\
k_{13}(u) &=&\frac{1}{{\cal D}_{2}(u,\eta )}\beta _{13}^{2}\cosh \frac{1}{2}%
\eta \cosh \frac{3}{2}\eta \cosh (u+\frac{1}{2}\eta )\sinh \eta \sinh 2u, 
\nonumber \\
k_{13}(u) &=&-\frac{\beta _{21}^{2}}{\beta _{12}^{2}}k_{13}(u),  \nonumber \\
k_{11}(u) &=&-\frac{1}{{\cal D}_{2}(u,\eta )}\left\{ \frac{1}{2}{\rm e}^{-%
\frac{2}{3}u}\cosh (u+\frac{1}{2}\eta )({\rm e}^{2u}\cosh 2\eta +\cosh \eta )%
{\cal F}_{2}(u,\eta )\right.  \nonumber \\
&&-\left. \beta _{12}^{2}{\rm e}^{-\frac{2}{3}u+2\eta }\cosh ^{3}\frac{1}{2}%
\eta \cosh \frac{3}{2}\eta \cosh (u+\frac{3}{2}\eta )\right\} ,  \nonumber \\
&&  \nonumber \\
k_{33}(u) &=&-\frac{1}{{\cal D}_{2}(u,\eta )}\left\{ \frac{1}{2}{\rm e}^{%
\frac{2}{3}u}\cosh (u+\frac{1}{2}\eta )({\rm e}^{-2u}\cosh 2\eta +\cosh \eta
){\cal F}_{2}(u,\eta )\right.  \nonumber \\
&&-\left. \beta _{23}^{2}{\rm e}^{\frac{2}{3}u-2\eta }\cosh ^{3}\frac{1}{2}%
\eta \cosh \frac{3}{2}\eta \cosh (u+\frac{3}{2}\eta )\right\} ,
\label{eq6.12}
\end{eqnarray}
where 
\begin{eqnarray}
{\cal D}_{2}(u,\eta ) &=&\beta _{12}^{2}{\rm e}^{2\eta }\cosh ^{3}\frac{1}{2}%
\eta \cosh \frac{3}{2}\eta \cosh (u-\frac{3}{2}\eta )  \nonumber \\
&&-\cosh (u-\frac{1}{2}\eta )\cosh (u+\frac{1}{2}\eta )\cosh (u+\frac{3}{2}%
\eta ){\cal F}_{2}(u,\eta ),  \label{eq6.14}
\end{eqnarray}
and 
\begin{equation}
{\cal F}_{2}(u,\eta )=\beta _{12}^{2}{\rm e}^{2\eta }\cosh \frac{1}{2}\eta
\cosh \frac{3}{2}\eta -2\beta _{13}\sinh \eta .  \label{eq6.15}
\end{equation}

\subsection{ Type-II Solution}

The third solution is more simple. Now the parameters are related by 
\begin{eqnarray}
\beta _{33} &=&-{\rm e}^{-2\eta }\beta _{11}-\frac{8}{3}{\rm e}^{-\eta
}\sinh \eta ,  \nonumber \\
\beta _{31} &=&-\frac{1}{\beta _{13}}{\rm e}^{-2\eta }(\beta _{11}-\frac{4}{3%
})^{2},  \label{eq6.27}
\end{eqnarray}
and the corresponding $K$-matrix has the form (\ref{eq2.29}) with the
following non-zero entries 
\begin{eqnarray}
k_{31}(u) &=&\frac{\beta _{31}}{\beta _{13}}k_{13}(u),  \nonumber \\
k_{13}(u) &=&\frac{1}{2}\beta _{13}\frac{\sinh 2u}{1-(\beta _{11}-\frac{4}{3}%
){\rm e}^{-\eta }\sinh u\sinh (u+\eta )},  \nonumber \\
k_{11}(u) &=&{\rm e}^{-\frac{2}{3}u}\frac{1+2e^{u}\sinh u+(\beta _{11}-\frac{%
4}{3}){\rm e}^{u-\eta }\cosh \eta \sinh u}{1-(\beta _{11}-\frac{4}{3}){\rm e}%
^{-\eta }\sinh u\sinh (u+\eta )},  \nonumber \\
k_{33}(u) &=&{\rm e}^{\frac{2}{3}u}\frac{1-2e^{-u}\sinh u-(\beta _{11}-\frac{%
4}{3}){\rm e}^{-u-\eta }\cosh \eta \sinh u}{1-(\beta _{11}-\frac{4}{3}){\rm e%
}^{-\eta }\sinh u\sinh (u+\eta )}.  \label{eq6.28}
\end{eqnarray}
The procedure to calculate (\ref{eq6.28}) is the same ones used to calculate
the type-II solution of the {\small IK} model. In the appendix D we present
all reduced solutions for the $osp(2|1)$ model.

\section{Boundary Integrable Hamiltonians}

In order to derive the Hamiltonians it is convenient to expand the $R$%
-matrix around the regular point $u=0$. For $19$-vertex models the
corresponding solutions with the standard normalization can be read directly
from (\ref{eq2.1}). They have the form 
\begin{equation}
PR(u,\eta )=1+u(\alpha ^{-1}h+\beta )+{\large O}(u^{2}).  \label{eq7.1}
\end{equation}
with $\alpha $ and $\beta $ being scalar functions.

The two-site Hamiltonian $H_{k,k+1}$ (\ref{int.5}) is the $h$ operator in (%
\ref{eq7.1}) acting on the quantum spaces at sites $k$ and $k+1$. From (\ref
{eq7.1}) we obtain 
\begin{equation}
H_{k,k+1}=\left( 
\begin{array}{lllllllll}
z_{1} & 0 & \ 0 & 0 & \ 0 & 0 & \ 0 & 0 & 0 \\ 
0 & \stackrel{\_}{z}_{5} & \ 0 & 1 & \ 0 & 0 & \ 0 & 0 & 0 \\ 
0 & 0 & \ \stackrel{\_}{z}_{7} & 0 & \ \stackrel{\_}{z}_{6} & 0 & \ z_{3} & 0
& 0 \\ 
0 & 1 & \ 0 & z_{5} & \ 0 & 0 & 0 & 0 & 0 \\ 
0 & 0 & \varepsilon \stackrel{\_}{z}_{6} & 0 & \varepsilon z_{4} & 0 & 
\varepsilon z_{6} & 0 & 0 \\ 
0 & 0 & \ 0 & 0 & \ 0 & \stackrel{\_}{z}_{5} & \ 0 & 1 & 0 \\ 
0 & 0 & \ z_{3} & 0 & \ z_{6} & 0 & \ z_{7} & 0 & 0 \\ 
0 & 0 & \ 0 & 0 & \ 0 & 1 & \ 0 & z_{5} & 0 \\ 
0 & 0 & \ 0 & 0 & \ 0 & 0 & \ 0 & 0 & z_{1}
\end{array}
\right) ,
\end{equation}
which can be easily written in terms of the usual spin-$1$ operators: 
\begin{eqnarray}
H_{k,k+1} &=&\epsilon z_{4}+\frac{1}{2}(\stackrel{\_}{z}%
_{5}-z_{5})[S_{k}^{z}-S_{k+1}^{z}]+\frac{1}{2}(z_{5}+\stackrel{\_}{z}%
_{5}-2\epsilon z_{4})[(S_{k}^{z})^{2}+(S_{k+1}^{z})^{2}]  \nonumber \\
&&+\frac{1}{4}(2z_{1}-z_{7}-\stackrel{\_}{z}_{7})S_{k}^{z}S_{k+1}^{z}+\frac{1%
}{4}[2(z_{1}-z_{5}-\stackrel{\_}{z}_{5})+x_{7}+\stackrel{\_}{z}%
_{7}+4\epsilon z_{4}](S_{k}^{z}S_{k+1}^{z})^{2}  \nonumber \\
&&+\frac{1}{4}(z_{7}-\stackrel{\_}{z}_{7}-2(x_{5}-\stackrel{\_}{z}%
_{5}))[(S_{k}^{z})^{2}S_{k+1}^{z}-S_{k}^{z}(S_{k+1}^{z})^{2}]  \nonumber \\
&&+\frac{1}{4}z_{3}[(S_{k}^{+}S_{k+1}^{-})^{2}+(S_{k}^{-}S_{k+1}^{+})^{2}]-%
\frac{1}{2}[\epsilon z_{6}S_{k}^{+}S_{k+1}^{-}+\epsilon \stackrel{\_}{z}%
_{6}S_{k}^{-}S_{k+1}^{+}]S_{k}^{z}S_{k+1}^{z}  \nonumber \\
&&-\frac{1}{2}S_{k}^{z}S_{k+1}^{z}[\stackrel{\_}{z}%
_{6}S_{k}^{+}S_{k+1}^{-}+z_{6}S_{k}^{-}S_{k+1}^{+}]+\frac{1}{2}\left\{
S_{k}^{+}S_{k}^{z}S_{k+1}^{z}S_{k+1}^{-}+S_{k}^{-}S_{k}^{z}S_{k+1}^{z}S_{k+1}^{+}\right.
\nonumber \\
&&\left.
+S_{k}^{z}S_{k}^{+}S_{k+1}^{-}S_{k+1}^{z}+S_{k}^{z}S_{k}^{-}S_{k+1}^{+}S_{k+1}^{z}\right\}
\label{eq7.2}
\end{eqnarray}
where 
\begin{eqnarray}
\varepsilon &=&1,\quad \alpha =\sinh 2\eta ,\quad \beta =0,\quad
z_{1}=0,\quad z_{3}=-1,\quad z_{4}=-2\cosh 2\eta ,  \nonumber \\
z_{5} &=&\stackrel{\_}{z}_{5}=-\cosh 2\eta ,\quad z_{6}=\stackrel{\_}{z}%
_{6}=2\cosh \eta ,\quad z_{7}=\stackrel{\_}{z}_{7}=-1-2\cosh 2\eta
\label{eq7.3}
\end{eqnarray}
for the {\small ZF} model; 
\begin{eqnarray}
\varepsilon &=&-1,\quad \alpha =\sinh 2\eta ,\quad \beta =0\qquad
z_{1}=0,\quad z_{3}=1,\quad z_{4}=2\cosh 2\eta ,  \nonumber \\
\stackrel{\_}{z}_{5} &=&z_{5}=-\cosh 2\eta ,\quad \stackrel{\_}{z}%
_{6}=z_{6}=2\sinh \eta ,\quad \stackrel{\_}{z}_{7}=z_{7}=-1-4\sinh ^{2}\eta
\label{eq7.4}
\end{eqnarray}
for the $sl(2|1)$ model; 
\begin{eqnarray}
\varepsilon &=&1,\ \alpha =-2\sinh 2\eta ,\ \beta =0,\ z_{1}=0,\ z_{3}=\frac{%
\cosh \eta }{\cosh 3\eta },\ z_{4}=-2\frac{\sinh 4\eta \sinh \eta }{\cosh
3\eta },  \nonumber \\
z_{5} &=&-{\rm e}^{-2\eta },\ \stackrel{\_}{z}_{5}=-{\rm e}^{2\eta },\ z_{6}=%
{\rm e}^{2\eta }\frac{\sinh 2\eta }{\cosh 3\eta },\ \stackrel{\_}{z}_{6}=-%
{\rm e}^{-2\eta }\frac{\sinh 2\eta }{\cosh 3\eta }  \nonumber \\
z_{7} &=&-{\rm e}^{-4\eta }\frac{\cosh \eta }{\cosh 3\eta },\ \stackrel{\_}{z%
}_{7}=-{\rm e}^{4\eta }\frac{\cosh \eta }{\cosh 3\eta }  \label{eq7.5}
\end{eqnarray}
for the {\small IK} model and 
\begin{eqnarray}
\varepsilon &=&-1,\ \alpha =\sinh 2\eta ,\ \beta =-\coth 2\eta ,\
z_{1}=\cosh 2\eta ,\ z_{3}=\frac{\sinh \eta }{\sinh 3\eta },  \nonumber \\
\ z_{4} &=&1+\coth 3\eta \sinh 2\eta ,\ z_{5}=-\frac{\sinh 2\eta }{3},\ 
\stackrel{\_}{z}_{5}=-z_{5},\ z_{6}=-{\rm e}^{-2\eta }\frac{\sinh 2\eta }{%
\sinh 3\eta },  \nonumber \\
\ \stackrel{\_}{z}_{6} &=&{\rm e}^{2\eta }\frac{\sinh 2\eta }{\sinh 3\eta }%
,\ z_{7}=\frac{\sinh 2\eta }{3}+{\rm e}^{-\eta }\frac{\sinh 2\eta }{\sinh
3\eta },\ \stackrel{\_}{z}_{7}=-\frac{\sinh 2\eta }{3}+{\rm e}^{\eta }\frac{%
\sinh 2\eta }{\sinh 3\eta }  \label{eq7.6}
\end{eqnarray}
for the $osp(2|1)$ model.

Next, we recall (\ref{int.6}) to derive the boundary terms which can now be
read from the $K$-matrices obtained in this paper. Due to (\ref{eq2.4}), the
first boundary term is easily obtained: 
\begin{equation}
\left. \frac{dK_{-}(u)}{du}\right| _{u=0}=\left( 
\begin{array}{ccc}
\beta _{11} & \beta _{12} & \beta _{13} \\ 
\beta _{21} & 0 & \beta _{23} \\ 
\beta _{31} & \beta _{32} & \beta _{33}
\end{array}
\right)
\end{equation}
where $\beta _{ij}$ are the parameters calculated in the previous sections
for the general solutions of the $19$-vertex models considered in this
paper. Therefore the integrable\ spin-$1$ Hamiltonians with general boundary
interactions associated with these vertex models can be write in the form 
\begin{equation}
H=\sum_{k=1}^{N-1}H_{k,k+1}+\frac{1}{2}\sum_{i,j=1}^{3}\beta _{ij}{\rm E}%
_{ij}^{(1)}\otimes {\rm 1}+\frac{1}{2}\sum_{i,j=1}^{3}\alpha _{ij}{\rm 1}%
\otimes {\rm E}_{ij}^{(N)}  \label{eq7.7}
\end{equation}
where ${\rm E}_{ij}$ is an $3\times 3$ matrix with only non-vanishing entry $%
1$ in row $i$ and column $j$. $\alpha _{ij}$ are new parameters associated
with the left $K$-matrices, obtained from the second boundary term ${\rm tr}%
_{0}\stackrel{0}{K}_{+}(0)H_{N,0}/{\rm tr}K_{+}(0)$ through the
correspondence (\ref{int.8}), {\it i.e}. 
\begin{equation}
K_{-}(u,\beta _{ij})\longrightarrow K_{+}(u,\alpha _{ji})=K_{-}^{t}(-u-\rho
,\alpha _{ji})M  \label{eq7.8}
\end{equation}
For the graded models the same results follow after we use the graded
formulation for (\ref{eq7.7}) and (\ref{eq7.8}).

\section{Conclusion}

We have considered the boundary {\small YB} equation for some $19$-vertex
models. After a systematic study of the functional equations we find that
there is only a complete solution with three free parameters for the {\small %
ZF} and $sl(2|1)$ models. From vanishing of these parameters we can derive
some particular solutions, which we called reduced solutions. In these
models (see appendices A and B) the last possible reduction will give us a
diagonal solution with one free parameter.

For {\small IK} and $osp(2|1)$ models we find two complete solutions but
with only two free parameters. This decrease in the number of free
parameters is responsible for the appearance of a third solution, the
type-II solution, which also have two free parameters but with some
vanishing entries. From these general solutions we find reduced solutions by
vanishing of their free parameters (see appendices C and D). The last
possible reduction for these models gives us three diagonal solutions with
no free parameters. This explains the result obtained for the {\small IK}
model in a previous work \cite{Mezincescu4}.

\vspace{1.0cm}{}

{\bf Acknowledgment:} This work was supported in part by Funda\c{c}\~{a}o de
Amparo \`{a} Pesquisa do Estado de S\~{a}o Paulo--FAPESP--Brasil and by
Conselho Nacional de Desenvol\-{}vimento--CNPq--Brasil.

\vspace{1cm}{}

\centerline{\bf Appendix A.  ZF  Reduced Solutions} \setcounter{equation}{0} %
\renewcommand{\theequation}{A.\arabic{equation}}

In this appendix we will present particular solutions of the {\small RE}\
for the {\small ZF} model. These solutions are obtained by vanishing some
free parameters of the general solution derived in the section $3$.
Following the sub-classification discussed in the section $2$ we have

{\bf Case (i)} $\beta _{13}\neq 0$ and $\beta _{31}\neq 0$: The reduced
solution in this case was presented in the section $3$ . It is given by (\ref
{eq3.14}).

{\bf Case (ii)} $\beta _{13}\neq 0$ and $\beta _{31}=0$: In this case we
have two regular solutions 
\begin{equation}
K_{I_{b}}=\left( 
\begin{array}{ccc}
k_{11} & k_{12} & k_{13} \\ 
0 & 1 & k_{23} \\ 
0 & 0 & k_{33}
\end{array}
\right) \qquad {\rm and}\qquad K_{I_{c}}=\left( 
\begin{array}{ccc}
k_{11} & 0 & k_{13} \\ 
0 & 1 & 0 \\ 
0 & 0 & k_{33}
\end{array}
\right)  \label{A.1}
\end{equation}
The parameters for the solution $K_{I_{b}}$ are obtained by taking the limit 
$\beta _{21}\rightarrow 0$ in\ (\ref{eq3.6}). From this limit we have the
following parameters 
\begin{eqnarray}
\beta _{12} &=&-\frac{1}{2}\beta _{23}(\beta _{11}\sinh \eta -2\cosh \eta ) 
\nonumber \\
\beta _{33} &=&2\frac{\beta _{11}\cosh \eta -2\sinh \eta }{\beta _{11}\sinh
\eta -2\cosh \eta }  \nonumber \\
\beta _{13} &=&\frac{1}{8}\frac{\beta _{23}^{2}\sinh \eta (\beta _{11}\sinh
\eta -2\cosh \eta )}{\cosh \eta }  \nonumber \\
\beta _{11} &\neq &2\coth \eta  \label{A.2}
\end{eqnarray}
Substituting (A2) into (\ref{eq3.7}) we find 
\begin{eqnarray*}
k_{12}(u) &=&\frac{1}{2}\frac{\beta _{23}(\beta _{11}\sinh \eta -2\cosh \eta
)\sinh 2u}{\beta _{11}\sinh u-2\cosh (u)} \\
k_{23}(u) &=&-\frac{1}{2}\frac{\beta _{23}(\beta _{11}\sinh \eta -2\cosh
\eta )\sinh 2u}{\beta _{11}\sinh (u-\eta )+2\cosh (u-\eta )}
\end{eqnarray*}
\[
k_{13}(u)=-\frac{\beta _{13}(\beta _{11}\sinh \eta -2\cosh \eta )\sinh
2u\sinh (2u-\eta )}{\sinh \eta (\beta _{11}\sinh u\!-\!2\cosh u)(\beta
_{11}\sinh (u-\eta )\!+\!2\cosh (u-\eta ))} 
\]
\begin{eqnarray}
k_{11}(u) &=&-\frac{\beta _{11}\sinh u+2\cosh u}{\beta _{11}\sinh u-2\cosh u}
\nonumber \\
k_{33}(u) &=&-\frac{\beta _{11}\sinh (u+\eta )-2\cosh (u+\eta )}{\beta
_{11}\sinh (u-\eta )+2\cosh (u-\eta )}  \label{A.3}
\end{eqnarray}
Note that we have choose to write the parameters in terms of $\beta _{11}$
and $\beta _{23}$. The solution $K_{I_{c}}$ is the limit $\beta
_{23}\rightarrow 0$ of $K_{I_{b}}$. In this case there is no relation
between $\beta _{11}$ and $\beta _{13}$ but, using \ the equation $E[1,3]=0$
one can see that it is a two-parameter solution provided that 
\begin{equation}
\sinh 4\eta =0,\quad \sinh 2\eta \neq 0\quad {\rm and} \quad \beta_{11}\neq
2\coth \eta  \label{A.4}
\end{equation}

{\bf Case (iii) }$\beta _{13}=0$ and $\beta _{31}\neq 0$: This case is a
transposition of the previous one. The corresponding $K$-matrices are 
\begin{equation}
K_{I_{d}}=\left( 
\begin{array}{ccc}
k_{11} & 0 & 0 \\ 
k_{21} & 1 & 0 \\ 
k_{31} & k_{32} & k_{33}
\end{array}
\right) \qquad {\rm and}\qquad K_{I_{e}}=\left( 
\begin{array}{ccc}
k_{11} & 0 & 0 \\ 
0 & 1 & 0 \\ 
k_{31} & 0 & k_{33}
\end{array}
\right)  \label{A.5}
\end{equation}
Their parameters and non-zero entries are obtained from (\ref{A.2}) and\ (%
\ref{A.3}) respectively, using the interchange rule (\ref{eq2.7}). For $%
K_{I_{e}}$ we have the condition (\ref{A.4}) still.

{\bf Case (iv) }$\beta _{13}=0$ and $\beta _{31}=0$: \ Finally we arrive at
the diagonal solution. It has the form 
\begin{equation}
K_{I_{f}}=\left( 
\begin{array}{ccc}
k_{11} & 0 & 0 \\ 
0 & 1 & 0 \\ 
0 & 0 & k_{33}
\end{array}
\right)   \label{A.6}
\end{equation}
which is an one-parameter solution with 
\begin{eqnarray}
k_{11}(u) &=&-\frac{\beta _{11}\sinh u+2\cosh u}{\beta _{11}\sinh u-2\cosh u}%
,  \nonumber \\
k_{33}(u) &=&-\frac{\beta _{11}\sinh (u+\eta )-2\cosh (u+\eta )}{\beta
_{11}\sinh (u-\eta )+2\cosh (u-\eta )}.  \label{A.7}
\end{eqnarray}
It is obtained, for instance, from the solution $K_{I_{b}}$ when $\beta
_{23}\rightarrow 0$. This diagonal solution was derived in \cite{Mezincescu3}%
.

\newpage

\centerline{\bf Appendix B. IK Reduced Solutions} \setcounter{equation}{0} %
\renewcommand{\theequation}{B.\arabic{equation}}

For the {\small IK} model we have three general solutions. Therefore, we
will find three types of reduced solutions. Reduced solutions from the
type-I solutions discussed in the section $4$ are here obtained using the
same limit procedure presented in the {\small ZF} model

{\bf Case (i)} $\beta _{13}\neq 0$ and $\beta _{31}\neq 0$ we have two
one-parameter solutions 
\begin{equation}
K_{I_{a}}=\left( 
\begin{array}{ccc}
k_{11} & 0 & k_{13} \\ 
0 & 1 & 0 \\ 
k_{31} & 0 & k_{33}
\end{array}
\right)  \label{B.1}
\end{equation}
where 
\begin{eqnarray}
k_{11}(u) &=&\frac{\cosh 2\eta \mp i{\rm e}^{-u}\sinh \eta }{\cosh 2\eta \pm
i\sinh (u-\eta )},\ k_{33}(u)=\frac{\cosh 2\eta \mp i{\rm e}^{u}\sinh \eta }{%
\cosh 2\eta \pm i\sinh (u-\eta )}  \nonumber \\
k_{13}(u) &=&\frac{\beta _{13}(\cosh 2\eta \mp i\sinh \eta )\sinh u}{\cosh
2\eta \pm i\sinh (u-\eta )},\ k_{31}(u)=\frac{\beta _{31}}{\beta _{13}}%
k_{13}(u)  \label{B.2}
\end{eqnarray}
with the relation 
\begin{equation}
\beta _{13}\beta _{31}(\cosh 2\eta \mp i\sinh \eta )^{2}=-1  \label{B.3}
\end{equation}
Here we observe that these reduced solutions can be derived from the type-II
solution. Substituting 
\begin{equation}
\beta _{11}=\mp \frac{ie^{-\eta }}{\cosh 2\eta \mp i\sinh \eta }  \label{B.4}
\end{equation}
into (\ref{eq5.19}) we will get (\ref{B.2}).

{\bf Case (ii)} $\beta _{13}\neq 0$ and $\beta _{31}=0$: \ Here we have four
reduced solutions 
\begin{equation}
K_{I_{b}}=\left( 
\begin{array}{ccc}
k_{11} & k_{12} & k_{13} \\ 
0 & 1 & k_{23} \\ 
0 & 0 & k_{33}
\end{array}
\right) \qquad {\rm and}\qquad K_{I_{c}}=\left( 
\begin{array}{ccc}
k_{11} & 0 & k_{13} \\ 
0 & 1 & 0 \\ 
0 & 0 & k_{33}
\end{array}
\right)  \label{B.5}
\end{equation}
where 
\begin{eqnarray}
\beta _{12}^{2} &=&-2{\rm e}^{2\eta }\beta _{13}\frac{\cosh \eta }{\cosh
2\eta \mp i\sinh \eta }  \nonumber \\
\Gamma _{1}(u) &=&\beta _{12}{\rm e}^{-\frac{1}{2}u}\frac{\cosh (\frac{u}{2}%
-\eta )\pm i\sinh \frac{u}{2}}{\cosh (u-\eta )}  \nonumber \\
k_{12}(u) &=&\frac{1}{\beta _{13}}\Gamma _{1}(u)k_{13}(u),\ k_{23}(u)=\frac{%
\pm i{\rm e}^{u-2\eta }}{\beta _{13}}\Gamma _{1}(u)k_{13}(u)  \nonumber \\
k_{13}(u) &=&\beta _{13}\frac{\cosh 3\eta \sinh u}{\cosh \eta }\frac{\cosh (%
\frac{u}{2}-\eta )\mp i\sinh \frac{u}{2}}{\cosh (\frac{u}{2}+3\eta )\mp
i\sinh \frac{u}{2}}  \nonumber \\
k_{11}(u) &=&{\rm e}^{-u}\frac{\cosh (\frac{u}{2}-3\eta )\pm i\sinh \frac{u}{%
2}}{\cosh (\frac{u}{2}+3\eta )\mp i\sinh \frac{u}{2}}  \nonumber \\
k_{33}(u) &=&{\rm e}^{u}\frac{\cosh (\frac{u}{2}-3\eta )\pm i\sinh \frac{u}{2%
}}{\cosh (\frac{u}{2}+3\eta )\mp i\sinh \frac{u}{2}}  \label{B.6}
\end{eqnarray}
The solutions\ $K_{I_{c}}$ are one-parameter solutions provided that 
\begin{equation}
\sinh 2\eta =0\quad {\rm and}\quad \sinh \eta \neq 0.  \label{B.7}
\end{equation}

{\bf Case (iii) } $\beta _{13}=0$ and $\beta _{31}\neq 0$: In this case we
have four solutions 
\begin{equation}
K_{I_{d}}=\left( 
\begin{array}{ccc}
k_{11} & 0 & 0 \\ 
k_{21} & 1 & 0 \\ 
k_{31} & k_{32} & k_{33}
\end{array}
\right) \qquad {\rm and}\qquad K_{I_{e}}=\left( 
\begin{array}{ccc}
k_{11} & 0 & 0 \\ 
0 & 1 & 0 \\ 
k_{31} & 0 & k_{33}
\end{array}
\right)  \label{B.8}
\end{equation}
they are obtained from (\ref{B.6}) by the interchange $k_{ij}\leftrightarrow
k_{ji}$ and $\beta _{ij}\leftrightarrow \beta _{ji}$.

{\bf Case (iv)} $\beta _{13}=0$ and $\beta _{31}=0$: In this case we have
two diagonal solutions with no free parameters 
\begin{equation}
K_{I_{f}}=\left( 
\begin{array}{ccc}
k_{11} & 0 & 0 \\ 
0 & 1 & 0 \\ 
0 & 0 & k_{33}
\end{array}
\right)  \label{B.9}
\end{equation}
where 
\begin{equation}
k_{11}(u)={\rm e}^{-u}\frac{\cosh (\frac{u}{2}-3\eta )\pm i\sinh \frac{u}{2}%
}{\cosh (\frac{u}{2}+3\eta )\mp i\sinh \frac{u}{2}},\ k_{33}(u)={\rm e}^{u}%
\frac{\cosh (\frac{u}{2}-3\eta )\pm i\sinh \frac{u}{2}}{\cosh (\frac{u}{2}%
+3\eta )\mp i\sinh \frac{u}{2}}  \label{B.10}
\end{equation}
Moreover, from the type-II solution (\ref{eq5.19}) we have three more
reduced solutions. The two first are one-parameter solutions 
\begin{equation}
K_{II_{a}}=\left( 
\begin{array}{ccc}
1 & 0 & \beta _{13}\sinh u \\ 
0 & 1 & 0 \\ 
0 & 0 & 1
\end{array}
\right) \qquad {\rm and}\qquad K_{II_{b}}=\left( 
\begin{array}{ccc}
1 & 0 & 0 \\ 
0 & 1 & 0 \\ 
\beta _{31}\sinh u & 0 & 1
\end{array}
\right)  \label{B.11}
\end{equation}
and the last is the trivial solution 
\begin{equation}
K_{II_{c}}=\left( 
\begin{array}{ccc}
1 & 0 & 0 \\ 
0 & 1 & 0 \\ 
0 & 0 & 1
\end{array}
\right)  \label{B.12}
\end{equation}
They belong to the cases (ii), (iii) and (iv), respectively. The diagonal
solutions (\ref{B.9}) and (\ref{B.12}) were obtained for the first time by
Mezincescu and Nepomechie \cite{Mezincescu4}.

\newpage

\centerline{\bf Appendix C. $sl(2|1)$ Reduced Solutions} %
\setcounter{equation}{0} \renewcommand{\theequation}{C.\arabic{equation}}

In this appendix we present the reduced solution for the $sl(2|1)$ model.
They are obtained by the vanishing of some parameters in the general
solution derived in section $5$. The limit procedure is the same used for
the {\small ZF} model:

{\bf Case (i)} $\beta _{13}\neq 0$ and $\beta _{31}\neq 0$: The
corresponding $K$-matrix has the form 
\begin{equation}
K_{I_{a}}=\left( 
\begin{array}{ccc}
k_{11} & 0 & k_{13} \\ 
0 & 1 & 0 \\ 
k_{31} & 0 & k_{33}
\end{array}
\right)   \label{C.1}
\end{equation}
where 
\begin{eqnarray}
k_{13}(u) &=&\frac{1}{2}\frac{\beta _{13}\cosh \eta \sinh 2u}{\cosh (2u+\eta
)},\quad k_{31}(u)=-\frac{2\sinh 2u}{\beta _{13}\cosh \eta \cosh (2u+\eta )}
\nonumber \\
k_{11}(u) &=&k_{33}(u)=\frac{\cosh \eta }{\cosh (2u-\eta )}  \label{C.2}
\end{eqnarray}
It is a one-parameter solution of the graded {\small RE}.

{\bf Case (ii)} {\bf \ }$\beta _{13}\neq 0$ and $\beta _{31}=0$: Here we
have two reduced solutions with two free parameters 
\begin{equation}
K_{I_{b}}=\left( 
\begin{array}{ccc}
k_{11} & k_{12} & k_{13} \\ 
0 & 1 & k_{23} \\ 
0 & 0 & k_{33}
\end{array}
\right) \qquad {\rm and}\qquad K_{I_{c}}=\left( 
\begin{array}{ccc}
k_{11} & 0 & k_{13} \\ 
0 & 1 & 0 \\ 
0 & 0 & k_{33}
\end{array}
\right)  \label{C.3}
\end{equation}
where 
\begin{eqnarray}
k_{13}(u) &=&-\frac{\beta _{13}(\beta _{11}\cosh \eta -2\sinh \eta )\sinh
2u\cosh (2u-\eta )}{\cosh \eta (\beta _{11}\sinh u\!-\!2\cosh u)(\beta
_{11}\cosh (u-\eta )\!+\!2\sinh (u-\eta ))}  \nonumber \\
k_{12}(u) &=&-\frac{1}{2}\frac{\beta _{23}(\beta _{11}\cosh \eta -2\sinh
\eta )\sinh 2u}{\beta _{11}\sinh u-2\cosh (u)}  \nonumber \\
k_{23}(u) &=&\frac{1}{2}\frac{\beta _{23}(\beta _{11}\cosh \eta -2\sinh \eta
)\sinh 2u}{\beta _{11}\cosh (u-\eta )+2\sinh (u-\eta )}  \nonumber \\
k_{11}(u) &=&-\frac{\beta _{11}\sinh u+2\cosh u}{\beta _{11}\sinh u-2\cosh u}
\nonumber \\
k_{33}(u) &=&\frac{\beta _{11}\cosh (u+\eta )-2\sinh (u+\eta )}{\beta
_{11}\cosh (u-\eta )+2\sinh (u-\eta )}  \label{C.4}
\end{eqnarray}
with the following relations for the parameters 
\begin{equation}
\beta _{13}=-\frac{1}{8}\frac{\beta _{23}^{2}\cosh \eta (\beta _{11}\cosh
\eta -2\sinh \eta )}{\sinh \eta },\quad \beta _{11}\neq 2\tanh \eta
\label{C.5}
\end{equation}
The parameter $\eta $ in the solution $K_{I_{c}}$ is solution of the
equation: 
\begin{equation}
\sinh 4\eta =0\quad {\rm and}\quad \sinh 2\eta \neq 0.  \label{C.6}
\end{equation}

{\bf Case (iii)} {\bf \ }$\beta _{13}=0$ and $\beta _{31}\neq 0$: This is a
super-transposition of the previous case $\ $%
\begin{equation}
K_{I_{d}}=\left( 
\begin{array}{ccc}
k_{11} & 0 & 0 \\ 
k_{21} & 1 & 0 \\ 
k_{31} & k_{32} & k_{33}
\end{array}
\right) \qquad {\rm and}\qquad K_{I_{e}}=\left( 
\begin{array}{ccc}
k_{11} & 0 & 0 \\ 
0 & 1 & 0 \\ 
k_{31} & 0 & k_{33}
\end{array}
\right)  \label{C.7}
\end{equation}
with matrix elements and parameters obtained from (\ref{C.4}) and (\ref{C.5}%
) respectively, through the interchange $k_{ij}\leftrightarrow k_{ji}$ and $%
\beta _{ij}\leftrightarrow \beta _{ji}$.

{\bf Case (iv)} {\bf \ }$\beta _{13}=0$ and $\beta _{31}=0$: \ In yhis case
we have a one-parameter diagonal solution 
\begin{equation}
K_{I_{f}}=\left( 
\begin{array}{ccc}
k_{11} & 0 & 0 \\ 
0 & 1 & 0 \\ 
0 & 0 & k_{33}
\end{array}
\right)  \label{C.8}
\end{equation}
where 
\begin{eqnarray}
k_{11}(u) &=&-\frac{\beta _{11}\sinh u+2\cosh u}{\beta _{11}\sinh u-2\cosh u}
\nonumber \\
k_{33}(u) &=&\frac{\beta _{11}\cosh (u+\eta )-2\sinh (u+\eta )}{\beta
_{11}\cosh (u-\eta )+2\sinh (u-\eta )}  \label{C.9}
\end{eqnarray}
This diagonal solution was already known \cite{Mezincescu2}.

\newpage

\centerline{\bf Appendix D. $osp(2|1)$ Reduced Solutions} %
\setcounter{equation}{0} \renewcommand{\theequation}{D.\arabic{equation}}

In this appendix we will present the three types of reduced solutions for
the $osp(2|1)$ model

{\bf Case (i)} $\beta _{13}\neq 0$ and $\beta _{31}\neq 0$: In this case we
have two one-parameter solutions
\begin{equation}
K_{I_{a}}=\left( 
\begin{array}{ccc}
k_{11} & 0 & k_{13} \\ 
0 & 1 & 0 \\ 
k_{31} & 0 & k_{33}
\end{array}
\right)  \label{D.1}
\end{equation}

where 
\begin{eqnarray}
k_{11}(u) &=&\frac{1}{2}{\rm e}^{-\frac{2}{3}u}\frac{\cosh \eta -{\rm e}%
^{2u}\cosh 2\eta }{\sinh (u-\frac{1}{2}\eta )\sinh (u+\frac{3}{2}\eta )} 
\nonumber \\
k_{33}(u) &=&\frac{1}{2}{\rm e}^{\frac{2}{3}u}\frac{\cosh \eta -{\rm e}%
^{-2u}\cosh 2\eta }{\sinh (u-\frac{1}{2}\eta )\sinh (u+\frac{3}{2}\eta )} 
\nonumber \\
k_{13}(u) &=&-\frac{1}{2}\frac{\beta _{13}\sinh \frac{1}{2}\eta \sinh \frac{3%
}{2}\eta \sinh 2u}{\sinh (u-\frac{1}{2}\eta )\sinh (u+\frac{3}{2}\eta )} 
\nonumber \\
k_{31}(u) &=&\frac{1}{2}\frac{\sinh 2u}{\beta _{13}\sinh \frac{1}{2}\eta
\sinh \frac{3}{2}\eta \sinh (u-\frac{1}{2}\eta )\sinh (u+\frac{3}{2}\eta )}
\label{D.2}
\end{eqnarray}
for the first solution and 
\begin{eqnarray}
k_{11}(u) &=&\frac{1}{2}{\rm e}^{-\frac{2}{3}u}\frac{\cosh \eta +{\rm e}%
^{2u}\cosh 2\eta }{\cosh (u-\frac{1}{2}\eta )\cosh (u+\frac{3}{2}\eta )} 
\nonumber \\
k_{33}(u) &=&\frac{1}{2}{\rm e}^{\frac{2}{3}u}\frac{\cosh \eta +{\rm e}%
^{-2u}\cosh 2\eta }{\cosh (u-\frac{1}{2}\eta )\cosh (u+\frac{3}{2}\eta )} 
\nonumber \\
k_{13}(u) &=&\frac{1}{2}\frac{\beta _{13}\cosh \frac{1}{2}\eta \cosh \frac{3%
}{2}\eta \sinh 2u}{\cosh (u-\frac{1}{2}\eta )\cosh (u+\frac{3}{2}\eta )} 
\nonumber \\
k_{31}(u) &=&-\frac{1}{2}\frac{\sinh 2u}{\beta _{13}\cosh \frac{1}{2}\eta
\cosh \frac{3}{2}\eta \cosh (u-\frac{1}{2}\eta )\cosh (u+\frac{3}{2}\eta )}
\label{D.3}
\end{eqnarray}
for the second solution.

As in the {\small IK} model the solutions of this case can be obtained from
the type-II solution. Substituting 
\begin{equation}
\beta _{11}=\frac{2\cosh 2\eta +2\sinh \eta +{\rm e}^{\eta }}{3\sinh \frac{1%
}{2}\eta \sinh \frac{3}{2}\eta }\quad \ {\rm and\quad }\ \beta _{11}=\frac{%
2\cosh 2\eta -2\sinh \eta -{\rm e}^{\eta }}{3\cosh \frac{1}{2}\eta \cosh 
\frac{3}{2}\eta }  \label{D.4}
\end{equation}
into (\ref{eq6.28}) we will get (\ref{D.2}) and (\ref{D.3}), respectively

{\bf Case (ii) } $\beta _{13}\neq 0$ and $\beta _{31}=0$ : Here we also have
one-parameter solutions 
\begin{equation}
K_{I_{b}}=\left( 
\begin{array}{ccc}
k_{11} & k_{12} & k_{13} \\ 
0 & 1 & k_{23} \\ 
0 & 0 & k_{33}
\end{array}
\right) \qquad {\rm and}\qquad K_{I_{c}}=\left( 
\begin{array}{ccc}
k_{11} & 0 & k_{13} \\ 
0 & 1 & 0 \\ 
0 & 0 & k_{33}
\end{array}
\right)  \label{D.5}
\end{equation}
with amplitudes gives by 
\begin{eqnarray}
\beta _{13} &=&\frac{1}{2}{\rm e}^{2\eta }\beta _{12}^{2}\frac{\sinh \frac{1%
}{2}\eta \sinh \frac{3}{2}\eta }{\sinh \eta }  \nonumber \\
k_{12}(u) &=&-\frac{1}{2}{\rm e}^{-\frac{1}{3}u}\beta _{12}\frac{\sinh \frac{%
3}{2}\eta }{\sinh (u-\frac{3}{2}\eta )}\sinh 2u  \nonumber \\
k_{23}(u) &=&-\frac{1}{2}{\rm e}^{\frac{1}{3}u+2\eta }\beta _{12}\frac{\sinh 
\frac{3}{2}\eta }{\sinh (u-\frac{3}{2}\eta )}\sinh 2u  \nonumber \\
k_{13}(u) &=&-\frac{1}{2}\beta _{13}\frac{\sinh \frac{3}{2}\eta }{\sinh 
\frac{1}{2}\eta }\frac{\sinh (u+\frac{1}{2}\eta )}{\sinh (u-\frac{3}{2}\eta )%
}\sinh 2u  \nonumber \\
k_{11}(u) &=&-{\rm e}^{-\frac{2}{3}u}\frac{\sinh (u+\frac{3}{2}\eta )}{\sinh
(u-\frac{3}{2}\eta )},\quad k_{33}(u)=-{\rm e}^{\frac{2}{3}u}\frac{\sinh (u+%
\frac{3}{2}\eta )}{\sinh (u-\frac{3}{2}\eta )}  \label{D.6}
\end{eqnarray}
for the first solution and 
\begin{eqnarray}
\beta _{13} &=&\frac{1}{2}{\rm e}^{2\eta }\beta _{12}^{2}\frac{\cosh \frac{1%
}{2}\eta \cosh \frac{3}{2}\eta }{\sinh \eta }  \nonumber \\
k_{12}(u) &=&\frac{1}{2}{\rm e}^{-\frac{1}{3}u}\beta _{12}\frac{\cosh \frac{3%
}{2}\eta }{\cosh (u-\frac{3}{2}\eta )}\sinh 2u  \nonumber \\
k_{23}(u) &=&\frac{1}{2}{\rm e}^{\frac{1}{3}u+2\eta }\beta _{12}\frac{\cosh 
\frac{3}{2}\eta }{\cosh (u-\frac{3}{2}\eta )}\sinh 2u  \nonumber \\
k_{13}(u) &=&-\frac{1}{2}\beta _{13}\frac{\cosh \frac{3}{2}\eta }{\cosh 
\frac{1}{2}\eta }\frac{\cosh (u+\frac{1}{2}\eta )}{\cosh (u-\frac{3}{2}\eta )%
}\sinh 2u  \nonumber \\
k_{11}(u) &=&{\rm e}^{-\frac{2}{3}u}\frac{\cosh (u+\frac{3}{2}\eta )}{\cosh
(u-\frac{3}{2}\eta )},\quad k_{33}(u)={\rm e}^{\frac{2}{3}u}\frac{\cosh (u+%
\frac{3}{2}\eta )}{\cosh (u-\frac{3}{2}\eta )}  \label{D.7}
\end{eqnarray}
for the second solution.

{\bf Case (ii)} $\beta _{13}=0$ and $\beta _{31}\neq 0$: Here we have the $K$%
-matrices 
\begin{equation}
K_{I_{d}}=\left( 
\begin{array}{ccc}
k_{11} & 0 & 0 \\ 
k_{21} & 1 & 0 \\ 
k_{31} & k_{32} & k_{33}
\end{array}
\right) \qquad {\rm and}\qquad K_{I_{e}}=\left( 
\begin{array}{ccc}
k_{11} & 0 & 0 \\ 
0 & 1 & 0 \\ 
k_{31} & 0 & k_{33}
\end{array}
\right)  \label{D.8}
\end{equation}
whose matrix elements are obtained from (\ref{D.6}) and (\ref{D.7}) by the
interchange $k_{ij}\leftrightarrow k_{ji}$ and $\beta _{ij}\leftrightarrow
\beta _{ji}$. For both $K_{I_{c}}$ and $K_{I_{e}}$ we have the condition 
\begin{equation}
\sinh 2\eta =0\qquad {\rm and}\qquad \sinh \eta \neq 0  \label{D.9}
\end{equation}

{\bf Case (iv)} $\beta _{13}=0$ and $\beta _{31}=0$: Here we have two
diagonal solutions with no free parameters 
\begin{equation}
K_{I_{f}}=\left( 
\begin{array}{ccc}
k_{11} & 0 & 0 \\ 
0 & 1 & 0 \\ 
0 & 0 & k_{33}
\end{array}
\right)  \label{D.10}
\end{equation}
where 
\begin{equation}
k_{11}(u)=-{\rm e}^{-\frac{2}{3}u}\frac{\sinh (u+\frac{3}{2}\eta )}{\sinh (u-%
\frac{3}{2}\eta )},\quad k_{33}(u)=-{\rm e}^{\frac{2}{3}u}\frac{\sinh (u+%
\frac{3}{2}\eta )}{\sinh (u-\frac{3}{2}\eta )}  \label{D.11}
\end{equation}
for the first solution and 
\begin{equation}
k_{11}(u)={\rm e}^{-\frac{2}{3}u}\frac{\cosh (u+\frac{3}{2}\eta )}{\cosh (u-%
\frac{3}{2}\eta )},\quad k_{33}(u)={\rm e}^{\frac{2}{3}u}\frac{\cosh (u+%
\frac{3}{2}\eta )}{\cosh (u-\frac{3}{2}\eta )}  \label{D.12}
\end{equation}
for the second solution.

In addition, we have three reduced solutions from the type-II solution. The
two first are one-parameter solutions 
\begin{equation}
K_{II_{a}}=\left( 
\begin{array}{ccc}
e^{\frac{4}{3}u} & 0 & \frac{1}{2}\beta _{13}\sinh 2u \\ 
0 & 1 & 0 \\ 
0 & 0 & e^{-\frac{4}{3}u}
\end{array}
\right) ,\quad K_{II_{b}}=\left( 
\begin{array}{ccc}
e^{\frac{4}{3}u} & 0 & 0 \\ 
0 & 1 & 0 \\ 
\frac{1}{2}\beta _{31}\sinh 2u & 0 & e^{-\frac{4}{3}u}
\end{array}
\right)  \label{D.13}
\end{equation}
and the last is a diagonal solution with no free parameter 
\begin{equation}
K_{II_{c}}=\left( 
\begin{array}{ccc}
e^{\frac{4}{3}u} & 0 & 0 \\ 
0 & 1 & 0 \\ 
0 & 0 & e^{-\frac{4}{3}u}
\end{array}
\right)  \label{D.14}
\end{equation}
They belong to the cases (ii), (iii) and (iv), respectively.

\end{document}